\documentclass[lettersize,journal,peerreview]{IEEEtran}

\usepackage{amsmath,amsfonts}
\usepackage{algorithm}
\usepackage{array}
\usepackage{subcaption}
\usepackage{textcomp}
\usepackage{stfloats}
\usepackage{url}
\usepackage{verbatim}
\usepackage{graphicx}
\usepackage{cite}
\usepackage{algpseudocode}
\usepackage{hyperref}
\usepackage{listings}
\usepackage[table]{xcolor}

\begin{document}

\title{Programmable Data Plane Switch based Heavy Hitter Flow Detection using Packet-Count and Packet-Size features
} 

\author{
  \textsf{Adarsha K. Sasidhar$^1$, Krishna M. Sivalingam$^1$ (Fellow, IEEE) and 
Gauravdeep Shami$^2$ (Member, IEEE), Marc Lyonnais$^2$ (Senior Member, IEEE) and   Rodney Wilson$ ^2$} \\
{$^1$Department of Computer Science and Engineering, Indian Institute of
 Technology Madras, Chennai, India} \\
{$^2$External Research, Global Research and Development, Ciena Corporation, Ottawa, Canada} \\ 
Email: {\small\texttt{\{cs22s017@cse, skrishnam@cse\}.iitm.ac.in,
    \{gshami,mlyonnai,rwilson\}@ciena.com}} 
}







\maketitle

\begin{abstract}
In data networks carrying large numbers of flows, Heavy Hitters (HHs) or Elephant flows are the flows exceeding  pre-determined thresholds (e.g. number of packets or bytes) in a given time window. Such HH flows need to be handled differently in order to minimize their impact on other smaller flows. HH detection techniques have been shown to be more effective when implemented in programmable data plane switches. In recent work, it was shown that the inter-packet gap can be used to identify heavy hitters. Such schemes use a limited-size hash table for storing flow state information and using this for the detection. However, when hash collisions occur, it is possible that a valid HH flow in the table can be replaced by a non-HH flow resulting in missing detection of HH flows. To address this problem, this paper incorporates a flow's medium-term Packet Count (PC) feature. In order to limit the packet count field size in the hash table, counting is done only till hash collision occurs so as to reduce the range of values to be stored and thus, the required number of bits. Also, another flow's medium-term feature, Packet Size (PS) is incorporated independently. 
The proposed scheme has been implemented in the P4 language and tested on Intel Tofino hardware. Performance evaluation has been performed using CAIDA and MAWI-based real-life traffic traces. 
The results show that in several scenarios cases, we can significantly
reduce the False Negatives for HHs by using the packet count data
effectively and efficiently. 
\end{abstract}


\IEEEpeerreviewmaketitle

\section{Introduction} \label{sec:Intro}

In data networks, multiple flows exist simultaneously, where a flow is defined as a long-term data session between a source and a destination.  The term ``Heavy Hitters (HHs)'' has been used to denote long-standing flows with large traffic volumes \cite{HH:SOSR17}.  In Data Center Networks (DCNs), these flows are classified into two types and are referred to as ``elephant'' flows and mice flows \cite{MiceLongRatio}.
In order to improve network utilization and flow-level performance, it is necessary to classify flows based on their traffic volume.  This has applications in data mining, information retrieval, databases, security and so on. Several earlier works to detect HHs used flow counters (raw counters or some optimized data structures), which divide the network stream into a fixed number of time slots \cite{sketchBasedHH}. HHs are defined as flows consuming more than a fraction of link capacity in a particular interval. This may lead to inaccuracies since the traffic trends from earlier time slots were not considered. 

Many of the current flow classification schemes use counter-based approaches and accumulation-based approaches. In counter-based approaches for HH detection \cite{counterBasedHH}, part of the traffic information is recorded in a table where the flow's table location is obtained by hashing some of the flow's header fields. The hash tables are limited in size due to memory constraints. When there is a hash collision due to multiple flows mapping to the same table location, the smallest flow is replaced by the new flow. Here, when the memory space is limited, heavy flows may be mistakenly replaced by small ones. 
In accumulation-based approaches, a special data structure (e.g. Sketch \cite{sketchBasedHH}) is used. This hashes each flow to the corresponding memory entries in a table and records the accumulated information of all traffic in real time. It has lower memory consumption and faster update speed and detection. When multiple flows hash to the same memory entry (called a \textit{bucket}), the sketches resolve the hash collision by keeping the accumulated value of flow information. Due to the accumulation of traffic values over time, small flows may be incorrectly identified as heavy ones.
Both approaches require a large memory capacity to store flow information and also require fast feedback. This limits the statistical information size, resulting in identifying small flows as large ones, causing an overestimation problem \cite{ChainSketchTON23}.

\IEEEpubidadjcol

Some approaches have been proposed to use machine learning (ML)
techniques for HH detection (\cite{rkk}). These techniques require the
control plane's significant involvement, increasing the
switch-controller latency. Also, it requires additional training time
and the availability of datasets.  Implementing ML techniques on the
switch is difficult or not possible due to limited set of operations
available on the switch and thus require approximations.  In some recent
works, various flow state features such as packet size distribution,
byte count, and flow duration were considered for HH detection
\cite{HH:SOSR17, rkk}. The per-flow Inter-Packet Gap (IPG) metric was
used for HH detection in the research work titled `HH-IPG'
\cite{hh-ipg}. We consider HH-IPG as the baseline approach. In this
paper, we further improve HH detection performance by combining IPG and
packet counters.

In  HH-IPG \cite{hh-ipg}, a table \textit{T} with \textit{m} slots and an associated hash function $h()$ are used to obtain and store the weighted Inter-Packet Gap (IPG) of the flows. Each slot consists of four fields: flow-id, last noted weighted IPG ($\mathrm{IPG}_f^{\omega-1,i}$), last noted timestamp ($\mathrm{TS}_f^{l,i}$), a dimensionless metric for the flow to store flow’s throughput state ($\tau_f^i$). Due to the limited hash table size, hash collisions occur. When this happens, the weighted IPG will increase, and if its value is higher than a pre-defined threshold, the previous entry will be replaced by the new entry. Here, some HH flows may not be detected if only IPG metric is used.  
In order to reduce the number of false negatives and increase the accuracy, we propose to include the \textit{Packet Count (PC)} metric along with IPG. PC is defined as the number of packets belonging to each flow and is used only when hash collisions occur. The PC value will be reset to zero after each hash overwriting. The advantage of checking PC only during collision is that the smaller number of bits will suffice to hold the count value. 

This paper contributes to the flow classification problem in
Programmable Data Plane (PDP) devices. The key contributions of this
paper are: (i) developed three versions of the proposed IPG+PC approach;
(ii) detailed heavy hitter detection algorithm and P4 pipeline,
improving the ones proposed in \cite{hh-ipg}; (iii) implementation in
Intel Tofino \cite{tofino-intel} switch and a python based simulation
model; (iv) analysis with real-life traffic traces in the simulation
model; (v) comparison of our approach with the baseline work and the
double hashing; (vi) use of the packet count value for deciding heavy
hitters during hash collision rather than after regular time intervals
as in traditional approaches; (vii) analysis of the trade-off of dynamic
memory requirement versus algorithm performance by the proposed
approach. 
 
An implementation-based study using the MAWI20 dataset (\cite{MAWI20})
shows that considering PC till hash collision along with IPG reduces the
number of false negatives for the HHs by up to 12\%. Specifically, on
10:00~AM packet trace with a time window of 5~seconds, considering PC
and IPG reduced the false negatives by 12.35\% compared with using only
the IPG metric.  
Also, for this trace, the F1 score for the baseline approach was 0.84276, which is improved to 0.875843 in version 1, to 0.881888 in version 2, and to 0.912665 in version 3 of the proposed approach. Similarly, the study using the CAIDA19 dataset \cite{CAIDA19url} shows that considering PC along with IPG reduced false negatives by 4.18\% compared to using only the IPG metric.

Below are the further inclusions and analysis performed in addition to our previous work (\cite{OurANTSpaper}): (i) Section~\ref{Sec:SDN} provides a relavant background for our work; (ii) Section~\ref{Sec:Algo} provides the proposed algorithm; (iii) Section~\ref{Sec:OptReq} discusses finetuning algorithm to optimize the results as per requirements of the specific application; (iv) Section~\ref{Sec:diffHash}: analysis of the effect of hash function being used in the proposed algorithm; (v) Section~\ref{Sec:DSsize}: observation of dynamic memory requirements of the proposed approach;  (vi) Section~\ref{Sec:doubleHash}: comparison of our approach with double hashing; (vii) Section~\ref{Sec:moreDataset} analysis on CAIDA dataset;  (viii) Section~\ref{Sec:pSize}: another feature packet size is used in place of packet count feature to perform classification along with IPG.

\section{Background}
This section presents the relevant background and preceding research.

\subsection{SDN, PDP switches, P4 language, PISA} \label{Sec:SDN}

The router or switch is a network device that can forward packets from
one device to another. In traditional switches, the data plane refers to
the tasks related to receiving a message (e.g., forwarding process of
network packets), and the control plane refers to the actions that
control the data plane (e.g., routing process). The control plane
determines how the network should behave, while the data plane
implements that behavior on individual packets
\cite{SDNwebsite}. Traditional switches have hard-wired functions as
they are made of Application-Specific Integrated Circuits
(ASICs). Hence, these switches can be configured but are not
programmable in the field. To address this, a new controller-based
network type called Software Defined Networking (SDN) is introduced,
which introduced the idea of separation of the control plane and data
plane.


SDN is a network operational model that uses controllers that centralize
some network functions and give operators programmatic control over
their networks, making the networks' control plane programmable and
flexible. The OpenFlow standard was introduced with SDN to provide
switch abstraction by standardizing what a switch does based on commonly
used switches.  In OpenFlow, the control plane is separated from the
networking devices and implemented in a remote software-based
controller. OpenFlow provides a standard way for the controller to
communicate with the switches (e.g., to populate packet forwarding rules
on the switch). OpenFlow is implemented using ASICs and supports only a
fixed set of protocols. The data plane switches became less intelligent
and cheaper. Also, the controller-switch communication resulted in
increased latency.

In OpenFlow, the controller can be programmed to change the switch behavior. However, the data plane switches are not programmable; this led to the development of Programmable Data Plane (PDP) switches. PDP switches can perform some computations inside the data plane switch itself. This reduces the switch-controller latency as they do not entirely depend on the controller. In PDP switches, the  behavior of the switch (e.g., how to process packets) can be defined by network operators via software, unlike traditional switches where it is hardwired in underlying ASICs. It also allows rapid deployment as the necessary changes are made using software updates rather than designing new hardware.

PDP switches were developed on Protocol Independent Switch Architecture (PISA) (\cite{PISA}) whose components are a programmable parser, match action pipeline, a programmable deparser, ingress and egress pipelines. The programmable parser parses the headers and extracts the header values, which are passed to the ingress block to look up in the match table for a match. On a match, the associated set of actions is executed and the packet is sent to the egress block, which has another match-action pipeline. After egress, the packet is sent to the deparser. Deparser attaches all the headers to the packet payload and it sends packet to the egress port.

A high-level programming language like P4 is introduced to program the PDP switches, making it easier for network operators to add new functionality to existing routers and switches. P4 (Programming Protocol-Independent Packet Processors) is a data plane programming language that specifies how packets should be processed in the network \cite{p42014Sigcomm}. With P4, instead of acting as mere forwarding entities, the data plane in switches can process a packet and perform necessary action. 
P4 supports basic data types, arithmetic operations (addition, subtraction), bit-wise operations (left shift, right shift, and, or, not), logical operations (logical and, logical or), and ternary operation. P4 does not have looping statements and floating point numbers to ensure packet processing happens at the line rate. Similarly, conditional statements cannot be used in the action blocks of P4 language. The match-action pipeline is the only way to implement any logic.
 
A P4 target is the specific hardware or software platform that interprets and executes the P4 code to perform packet processing tasks. The P4 targets can be either software based (e.g., simulated environments like Behavioral Model v2 (BMv2), p4c-behavioral or software switches like Open vSwitch (OVS)) or hardware based (e.g., Barefoot Tofino and Tofino 2, NetFPGA SUME, Pensando Capri, AMD Pensando second generation ELBA, NVIDIA Mellanox Spectrum).
P4 target we used is Intel Tofino (\cite{tofino-intel}) which is the P4-programmable ethernet switch ASICs built using PISA.

\subsection{Limitations in Programmable Data Plane (PDP) switch implementations}
While PDP switches like Tofino offer better performance and flexibility for SDN and network automation, they have some limitations as they need to achieve line rate switching speed. Ideally, we expect flexibility with respect to what the switch can do in any PDP switch, giving the network operators control over how the switch behaves by writing P4 code and modifying the switch behavior. The traditional switches, which are made up of ASICs, were able to achieve line rate, but they offered no flexibility.  On the other hand, programmable switches such as Tofino, offer flexibility by providing programmatic control over the network achieving line rate speeds but have many limitations.

P4 does not have floating point numbers, and it does not supports loops, pointers, or dynamic memory allocation. The Tofino switch has a fixed number of match-action pipeline stages (hence, no loops), and it supports up to a specified number of states in the Finite State Machines (FSMs) supported in the ingress parser. 
The P4 language for Tofino supports only limited operations and reduced
domain specific instruction set. It does not support multiplications and
divisions for the register actions, because of which bit operations
supported in recent targets like Tofino have to be used to approximate
these operations. However, the comparison operations are limited to a
fixed number of bits. Also, there are access limitations to memory
registers; the register can be accessed once per packet lifetime in the
Tofino switch ASIC. To resolve this, packet resubmission can be used
along with the packet metadata to differentiate the actual packets from
the resubmitted ones. However, the packet resubmission may raise
concerns with respect to increased congestion and throughput. The Tofino
switch also has a limited stateful memory (MB’s of SRAM in Tofino). All
the logic from any proposed algorithms has to be implemented using a
match action pipeline.

PDP switches have been considered for solving different problems using
hardware-based algorithm implementations \cite{pHeavy2021, hh-ipg, marina,
  adaflow}.

\subsection{HH detection}
In data networks, network traffic flow consists of long lived flows or elephant flows or heavy hitters (HHs) and short lived flows or mice flows. The elephant flows are the flows exceeding the pre-determined threshold (wrt. number of packets or bytes) in a time window.
Also, In DCNs, the majority of flows are small, with only a few kilobytes (KB) in size. From the analysis of DCN flows, it was shown that 99\% of flows are smaller than 100~megabytes (MB). However, more than 90\% of bytes are in flows between 100~MB and 1 gigabyte (GB) \cite{hhSkew}. Even the observations from Peer-to-Peer (P2P) systems also show that the distribution of flow sizes is highly skewed, with less than 10\% of the end host Internet Protocol (IP) addresses contributing around 99\% of the total traffic volume \cite{P2P_hhRatio}. Hence, mice flows are numerous, where as elephant flows being less in quantity contribute to most of the traffic.

Elephant flows throttle the mice flows and other elephant flows by consuming a disproportionate amount of buffer and link capacity. This results in either a drop or delay in the mice packets. Despite this, the Equal Cost Multi-Path (ECMP) routing used in DCNs uses a five-tuple to hash and decide a route, and it treats the elephants and mice the same and does not consider the flow characteristics while routing. However, flow type must be predicted and handled differently. This could help balance the load, improve link utilization, and identify congestion and security attacks (e.g., Distributed Denial-of-Service (DDoS) detection). Many network management applications like accounting, network capacity planning (e.g., to improve link utilization), load balancing, caching, congestion control (by dynamically scheduling HHs), anomaly detection, etc, and networking measurements benefit from it.

In traditional network devices, HH detection algorithms run in the switch/router's control plane, not the data plane, as the computational power is higher in the control plane.
With the advent of the Programmable Data Plane (PDP) devices and the P4 (Programming Protocol-Independent Packet Processors) language, HH detection can be done on the data plane itself. One popular P4 target is the Intel Tofino \cite{tofino-intel} chip, which improves efficiency and performance of HH detection \cite{HH:SOSR17, HH:SOSR18}; as the data plane is faster, closer to packets and hence reduces the control plane latency.

\subsection{Inter-Packet Gap (IPG) metric} \label{subsec:HH}

The work presented in \cite{hh-ipg} (referred to as HH-IPG) proposes an algorithm and a P4 pipeline design using per-flow Inter Packet Gap (IPG) metric to detect Heavy Hitters (HH) entirely in the data plane. The IPG metric is considered by observing that the smaller the flow's IPG, the number of packets in the flow in a time window would be more and hence  heavier the flow and higher its throughput. Since the flow's features value were directly used in decision making, involvement of control plane for HH detection is much lower when compared with the algorithms that use ML based techniques such as \cite{FlowLens}.

The implementation is based on a hash table ($T$) with $m$ entries and a corresponding hash function ($h$). The hash function inputs are the following packet header fields: source IP address, destination IP address, source port and destination port. The output is the hash table slot where information regarding the given mapped flow is maintained. Each entry in the table stores the weighted IPG value for this flow along with its calculated throughput state (explained later). If the throughput is above a certain threshold, then a flow is classified as a heavy-hitter. Since the number of entries in $T$ is limited, it is possible to have hash collisions when two different flows map to the same slot in $T$. In such cases, a HH flow in the hash table might be replaced by a new flow which is not HH leading to a false negative. It is shown that failing to detect a HH flow can impact the behavior of network-control applications, such as load balancing in \cite{hh-ipg}. Hence, reducing the number of false negatives is an important requirement.
 
Based on preliminary studies of the HH-IPG scheme on some data sets (\cite{MAWI20, CAIDA19url}, we observed that some of the heavy hitters were indeed being replaced by newer flows. In order to avoid this problem, another flow feature, namely, the flow's packet count (PC) was also measured. For the MAWI20 dataset \cite{MAWI20} and using the simulation code of \cite{hh-ipg},  Table~\ref{tab:PC} presents a snapshot the packet count values for the flows stored in the table at a given time instant. The values represent the sequence of PC values printed whenever hash collision was observed. As seen, there are some flows which have high packet counts (the values shown in bold at Table~\ref{tab:PC}). Our hypothesis is that these flows with high PC values are potential HHs and should not be replaced in the hash table. This observation provided the motivation to consider the PC metric in addition to the IPG metric, till hash collision for improving HH detection performance. 
The next section presents the details of the proposed scheme. 

\begin{table}[!hbtp]
\centering
\caption{Packet Count values when Hash Collision occurs}
\label{tab:PC}
\def\sep{\unskip, &}
\begin{tabular}{r@{, }r@{, }r@{, }r@{, }r@{, }r@{, }r@{, }r@{, }}
\hline
    35 & 13 & 61 & 19 & \textbf{ 3073} & 13 & 9 & 29\\\hline
    7 & 22 & 15 & 19 & 32 & 13 & 17 & 7 \\\hline
    11 & 22 & 22 & 11 & 22 & 59 & 13 & 10 \\\hline
    22 & 5 & 13 & 14 & 22 & 13 & 17 & 16 \\\hline
    7 & \textbf{ 10926} & 21 & 93 & 2 & 15 & 22 & 2 \\\hline
    307 & 6 & 22 & 6 & 3 & 23 & 167 & 11 \\\hline
    .... \\\hline
    7 & 70 & \textbf{ 1326} & 7 & 15 & 22 & 22 & 5 \\\hline
    40 & \textbf{5074} & .... \\\hline
    38 & 5 & 21 & 11 & 27 & 83 & 22 & 5 \\\hline
    \textbf{ 1415} & 16 & 5 & 31 & 709 & 22 & 9 & 7 \\\hline
    22 & 19 & 15 & 24 & 22 & 15 & 383 & 13 \\\hline    
\end{tabular}
\end{table}

\section{Augmenting detection using Packet Count} \label{Sec:Imple}

The proposed approach is implemented by adding an extra field \textit{Packet Count(PC)} in the data structure used to keep track of flows in the HH-IPG paper \cite{hh-ipg}. 

\subsection{Hash Table structure}

Let $P = \{P_1, P_2,P_3, \ldots{}, P_N \}$ be a network stream with $N$ packets and let $F = \{f_1, f_2,f_3, \ldots{}, f_M \}$ be the set of network flows.
There are $M$ flows ($f_i)$ in the network in $P$. For each flow $f \in F$, the  throughput state is denoted by $\tau_f$. A pre-defined threshold $\tau_{th}$ is set by the network administrator. A flow is defined as a heavy hitter if  $\tau_f \geq \tau_{th}$.

\begin{figure}[hbtp]
\centering
\includegraphics[width=0.47\textwidth]{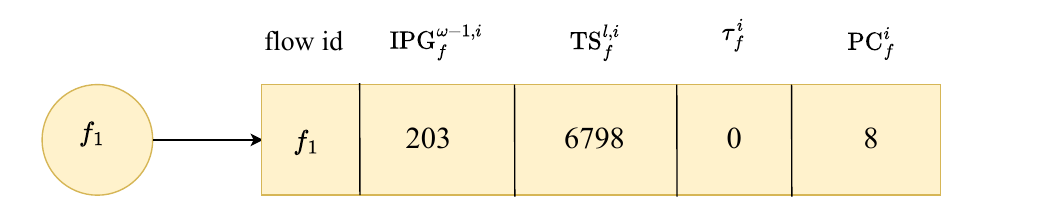}
 \caption{One example slot in the Data Structure.}
    \label{fig:DS}
\centering
\end{figure}

To store each flow's details, a hash table \textit{T} is used with \textit{m} slots and an associated hash function \textit{h(.)}, as shown in Fig.~\ref{fig:DS}. Each slot consists of five fields: flow id, last noted weighted IPG ($\mathrm{IPG}_f^{\omega-1,i}$), last noted timestamp ($\mathrm{TS}_f^{l,i}$), a dimensionless metric for the flow to store flow’s throughput state ($\tau_f^i$), and the Packet Count ($\mathrm{PC}_f^i$); where i $\in$ \{1,2,...,m\}. Here, $f$ indicates the flow to which the entry belongs, $i$ indicates the slot number. Also, in $\mathrm{IPG}_f^{\omega-1,i}$, $\omega-1$ means previous noted weighted IPG for considered slot and weight. 
To insert flow state values in \textit{T}, the corresponding flow identifier (id) is required. This is defined by four fields: source IP address, destination IP address, source port and destination port. The hash function \textit{h(.)} is applied on fid to determine the slot in $T$ for a given flow. Each time the existing flow is replaced by another flow during hash collision, the PC value will be reset to 1. The $\tau_f^i$ is used to decide whether the flow is HH or not. If $\tau_f^i \geq \tau_\mathrm{th}$ for a flow, then algorithm considers flow as HH. $\tau_\mathrm{th}$ is the threshold considered for the decision; threshold for IPG is considered to be $10,000\mu s$. The weighted IPG of flow $f$ (as defined in \cite{hh-ipg}) is computed as:

\begin{equation}
\mathrm{IPG}_f^{\omega} = \alpha \, \mathrm{IPG}_f^{\omega-1} + (1-\alpha) \, \mathrm{IPG}_f^c   
\label{eq:1}
\end{equation}

Here, $\mathrm{IPG}_f^{\omega-1}$ is the last noted weighted IPG, $\alpha \in [0,1]$ is the relative weight factor, and $\mathrm{IPG}_f^c$ is the current IPG which is the difference between last noted timestamp and the current timestamp. 
Also, small timeslot $T_{\omega t}$ is used to update $\tau_f^i$. The timestamp wraps around after every $T_{\omega t}$. More details about $T_{\omega t}$, $\tau_\mathrm{th}$, $\alpha$, and also how to choose them can be found in \cite{hh-ipg}. 

In the proposed approach, we use the flow's IPG feature for HH decision making as done in \cite{hh-ipg}, augmented with the packet counter (PC) feature. If $\tau_f^i \geq \tau_\mathrm{th}$ for a flow, then switch reports to controller that this flow is HH. In addition, at each hash collision, we check the packet counter (PC) value. If the PC is higher than the threshold $\mathrm{PC}_\mathrm{th}$, the given flow is also reported to the controller as HHs. In traditional packet counter based approaches, the HH decision is positive if the current PC value is greater than the threshold over some time duration or if this flow is among the Top-k PC valued flows. In this work, we use the PC value  when hash collision occurs.

\subsection{Hash Table insertion}

\begin{figure}[hbtp]
\centering
\includegraphics[width=0.47\textwidth]{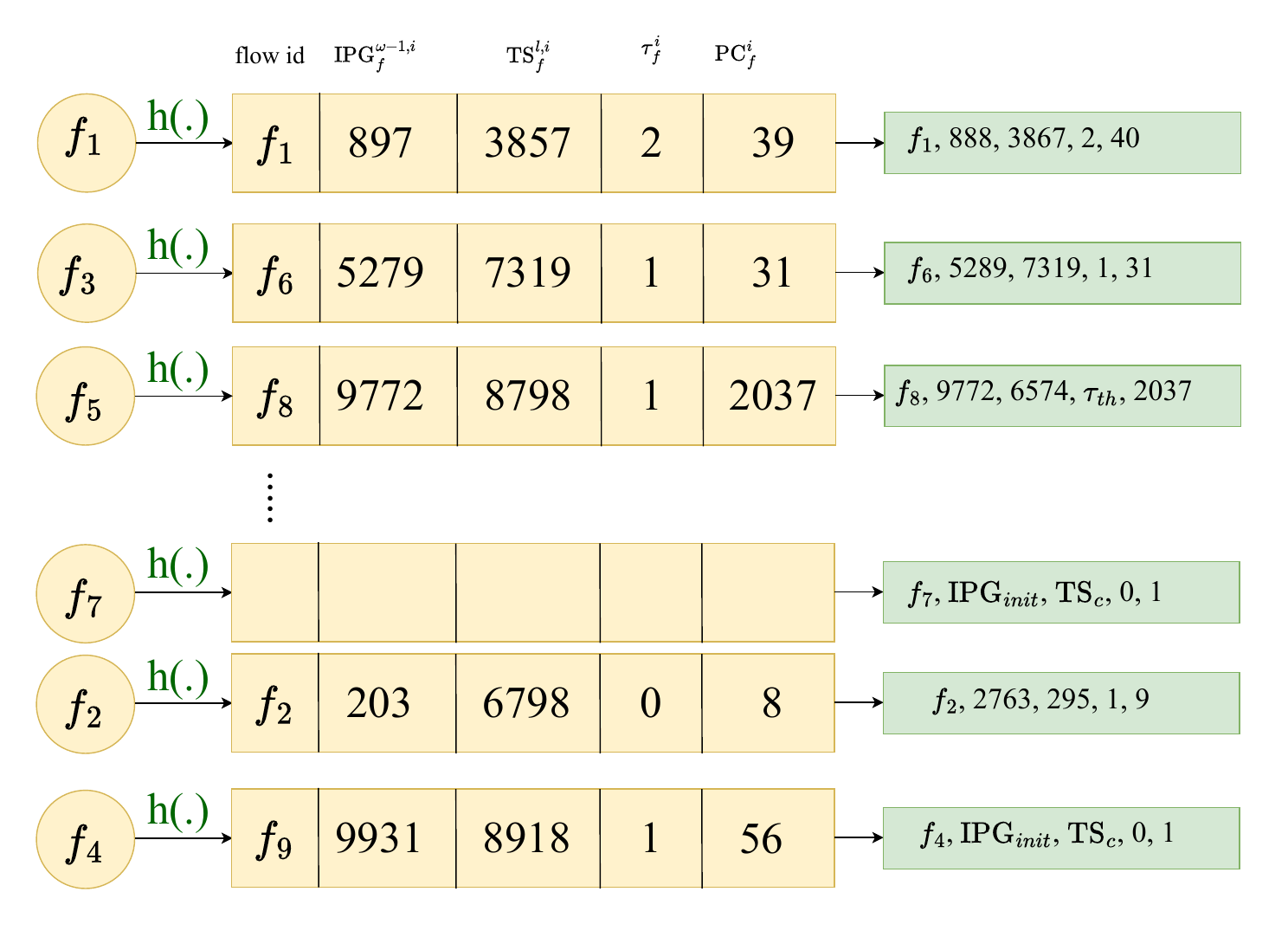}
 \caption{Hash table updates.}
    \label{fig:DS_Cases}
\centering
\end{figure}

During insertion into the hash table, there can be three possible cases for each incoming packet, as described below. An example of three possible cases is shown in Fig.~\ref{fig:DS_Cases}. These are suitably modified from the original scheme (\cite{hh-ipg}).

\textbf{Case 1: } The flow's hashed slot value is empty, implying that this is the first packet of the particular flow. The flow values (flow id of the current flow, $\mathrm{IPG}_\mathrm{init}$, $\mathrm{TS}_c$, 0, 1) are inserted in this slot. This initializes the flow id to hashed value of current packet's four fields: source IP, destination IP, source port and destination port. Next, $\mathrm{IPG}_f^{\omega-1,i}$ is set to $\mathrm{IPG}_\mathrm{init}$. Here,  $\mathrm{IPG}_\mathrm{init}$ is the initial IPG value which is equal to $\mathrm{IPG}_\mathrm{th}$ given by ($\mathrm{Packet Size}/\mathrm{HH}_\mathrm{th}$) \cite{hh-ipg}). Other fields set are: timestamp field to the current time stamp $\mathrm{TS}_c$,  $\tau_f^i$ to 0, and packet count to 1. 

For example, in Fig.~\ref{fig:DS_Cases}, when the first packet of the flow $f_7$ enters the programmable switch, the hash function is used to find its slot. Here it gets hashed to an empty slot. So we insert  $f_7$, $\mathrm{IPG}_\mathrm{init}$, $\mathrm{TS}_c$, 0, 1 into the slot to perform the specified initialization.


\textbf{Case 2: } There is an entry in the slot hashed to, with the same flow id. Here, we increment the packet count by 1, update timestamp, and calculate the IPG and update $\mathrm{IPG}_f^{\omega-1,i}$ based on this IPG as per Equation~\ref{eq:1}. Also, $\tau_f^i$ is updated by one, only when the current time stamp, $\mathrm{TS}_c < \mathrm{TS}_f^{l,i}$, where $\mathrm{TS}_f^{l,i}$ is the last noted timestamp. If this condition is not true, then $\tau_f^i$ is not changed.

For example, in Fig.~\ref{fig:DS_Cases}, when packet of $f_2$ arrives, the slot already has an entry with the same flow id. PC is incremented to 9 from 8. Since  $\mathrm{TS}_c < \mathrm{TS}_f^{l,i}$, it indicates that timestamp has been wrapped around. Hence, $\tau_f^i$ is updated with 1 and current IPG is calculated as $\mathrm{IPG}_f^c = \mathrm{TS}_c + T_\mathrm{wt} - \mathrm{TS}_f^{l,i}$. Then, $\mathrm{IPG}_f^{\omega-1,i}$ is calculated using this $\mathrm{IPG}_f^c$ in Equation~\ref{eq:1} and updated at the slot. In case of $f_1$, $\mathrm{TS}_c<\mathrm{TS}_f^{l,i}$ fails. So $\tau_f^i$ is not changed and $\mathrm{IPG}_f^c = \mathrm{TS}_c - \mathrm{TS}_f^{l,i}$. Equation~\ref{eq:1} gives $\mathrm{IPG}_f^{\omega-1,i}$ to be updated at the slot. Then, $\mathrm{PC}$ is incremented.

\textbf{Case 3:} There is already an entry in the hashed slot but with a different flow id. This is the case where hash collision occurs. Here, if $\mathrm{IPG}_f^{\omega-1,i} \leq \mathrm{IPG}_\mathrm{th}$, then $\mathrm{IPG}_f^{\omega-1,i}$ is linearly increased by adding a predefined and table size dependent constant $k$  \cite{hh-ipg}, and  the packet count is incremented by 1. If $\mathrm{IPG}_f^{\omega-1,i}>\mathrm{IPG}_\mathrm{th}$,  we check if the packet count of the existing entry exceeds the predefined threshold of PC. If yes, it will  be marked as a Heavy Hitter by changing $\tau_f$ value. Otherwise, the existing entry will be replaced by new entry.

In Fig.~\ref{fig:DS_Cases}, consider $f_3$, here as $\mathrm{IPG}_f^{\omega-1,i} \leq \mathrm{IPG}_\mathrm{th}$ holds, $\mathrm{IPG}_f^{\omega-1,i}$ is linearly increased by adding 10, as k=10 is considered. For $f_4$ and $f_5$,  since the condition $\mathrm{IPG}_f^{\omega-1,i} \leq \mathrm{IPG}_\mathrm{th}$ fails, condition $\mathrm{PC}_f^i > \mathrm{PC}_\mathrm{th}$ is tested for an existing entry. For $f_5$, since the PC exceeds the threshold, $f_8$ is marked as HH by changing $\tau_f^i$ value to $\tau_\mathrm{th}$. However, for $f_4$, $\mathrm{PC}_f^i > \mathrm{PC}_\mathrm{th}$ fails. Hence, $f_4$ will replace the existing entry.

\subsection{P4 implementation}

\begin{figure*}[hbtp]
\centering
\includegraphics[width=0.98\textwidth]{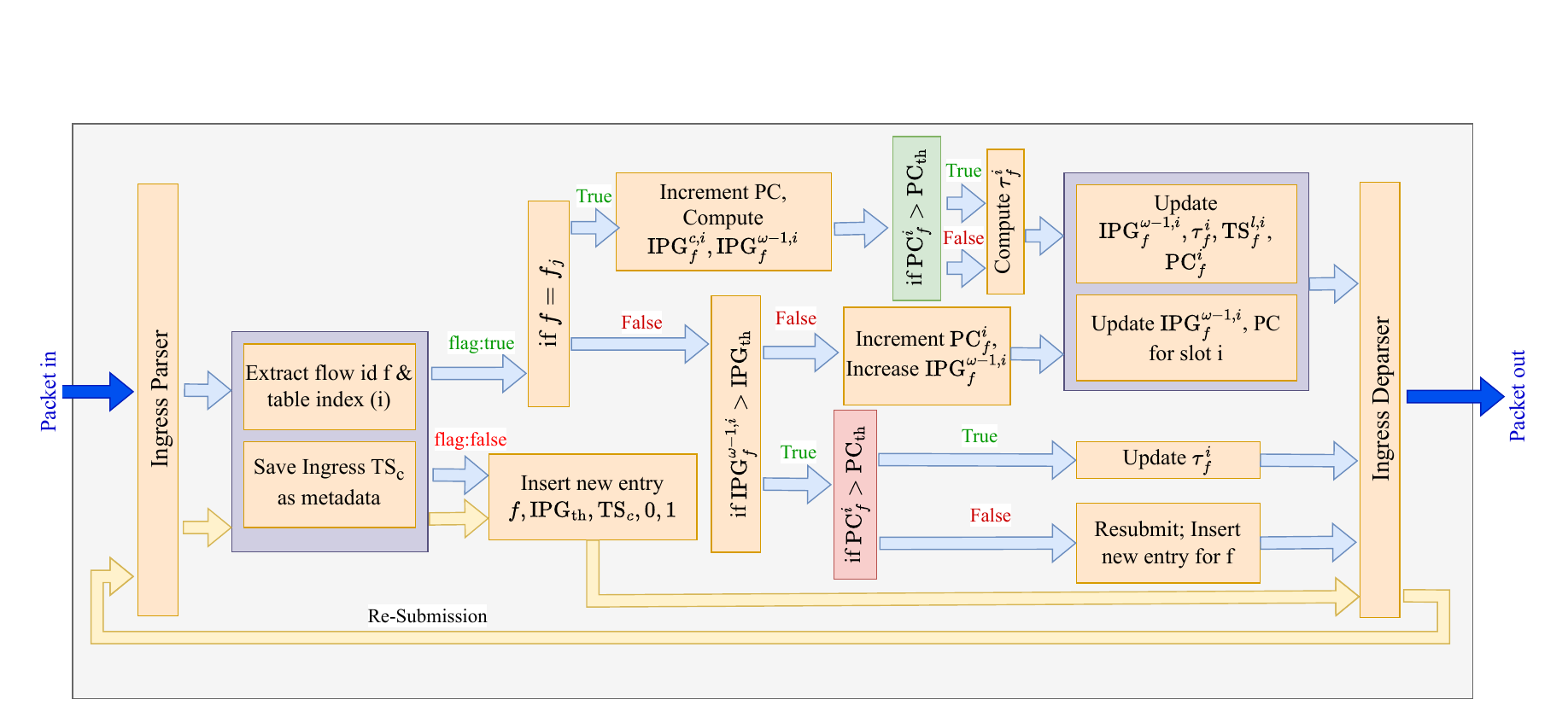}
 \caption{The proposed P4 pipeline for HH detection.}
    \label{fig:P4pipeline}
\centering
\end{figure*}

The proposed P4 pipeline implementation of the proposed HH detection scheme is shown in Fig.~\ref{fig:P4pipeline}. The pipeline design of \cite{hh-ipg} has been improved by adding the stages required for the inclusion of $\mathrm{PC}$. There is an access limitation to registers in the P4 ASIC. In a packet's lifetime, the register can be accessed only once. In the proposed pipeline, the register is accessed to check the flow ID initially. Hence, to re-access the register to replace the entry, packet re-submission is used \cite{hh-ipg}. Based on how the stored Packet Count (PC) value is used in decision making, three versions of the algorithm have been defined.

\textbf{Version 1 (V1):} During hash collision, if an existing current entry is being replaced, the packet count value of the existing flow in table is compared with packet count threshold. If the packet count value is greater, then the flow is considered as HH and reported to the controller. In Fig.~\ref{fig:P4pipeline}, this is shown by the red decision making box.

\textbf{Version 2 (V2):} For an existing flow (with no hash collision), it is classified as HH only if the flow's packet count has crossed the packet count threshold and IPG has crossed the IPG threshold. In Fig.~\ref{fig:P4pipeline}, this is shown by the green decision making box.

\textbf{Version 3 (V3):} This combines both the above versions, by considering both IPG and PC features. Thus, both collision and collision-less cases are considered.

\subsection{Proposed Algorithm} \label{Sec:Algo}
The proposed algorithm is presented in Algorithm~\ref{alg:HH-PC_IPG}. The algorithm is built by improving the one presented in \cite{hh-ipg}. Here, input $P_j$ is a packet where $1\leq j \leq N$ of flow $f$, $m$ is the total number of slots in the table \cite{hh-ipg}, $PC_f^i$ is packet count of flow $f$ in index $i$ of hash table.

\begin{algorithm}
    \caption{Proposed algorithm for HH detection using Inter Packet Gap and Packet Count.}
	\textbf{Input:} Packet $P_j$, $m$.
	\begin{algorithmic}[1]
    \State Get table index i from h(f), where i belongs to (1,2,...m).
    \If {flag=false}
        \State $flag \gets true$
        \State $f_i$, $\mathrm{IPG}_f^{\omega-1,i}$, $\mathrm{TS}_f^{l,i}$, $\tau_f^i$, $\mathrm{PC}_f^i$ = f, $\mathrm{IPG}_{init}$, $\mathrm{TS}_c$,0,1;
    \ElsIf{$f=f_i$}  
        \State $\mathrm{PC}_f^i = \mathrm{PC}_f^i+1$
        \If{$\mathrm{TS}_c > \mathrm{TS}_f^{l,i}$}
            \State $\mathrm{IPG}_f^c = \mathrm{TS}_c - \mathrm{TS}_f^{l,i};$
            \State $\mathrm{IPG}_f^{\omega-1,i} = \alpha.\mathrm{IPG}_f^{\omega-1,i} + (1-\alpha).\mathrm{IPG}_f^c;$
            \State $\mathrm{TS}_f^{l,i} = \mathrm{TS}_c$               
        \Else
            \State $\mathrm{IPG}_f^c = \mathrm{TS}_c + T_{wt} - \mathrm{TS}_f^{l,i};$
            \State $\mathrm{IPG}_f^{\omega-1,i} = \alpha.\mathrm{IPG}_f^{\omega-1,i} + (1-\alpha).\mathrm{IPG}_f^c;$
            \State $\mathrm{TS}_f^{l,i} = \mathrm{TS}_c$
            \State match on $\mathrm{IPG}_f^{\omega-1,i}$, set $metadata.tau$ as an action;
            \State $\tau_f^i = \tau_f^i + metadata.tau;$
        \EndIf
    \Else
        \If{$\mathrm{IPG}_f^{\omega-1,i} <= \mathrm{IPG}_{th}$}
            \State $\mathrm{IPG}_f^{\omega-1,i} = \mathrm{IPG}_f^{\omega-1,i} + k;$
        \Else
            \If{$\mathrm{PC}_f^i > \mathrm{PC}_{th}$}
                \State Set $\tau_f^i$ to value above threshold to mark flow as HH.
            \EndIf
            \State $f_i$, $\mathrm{IPG}_f^{\omega-1,i}$, $\mathrm{TS}_f^{l,i}$, $\tau_f^i$, $\mathrm{PC}_f^i$ = f, $\mathrm{IPG}_{init}$, $\mathrm{TS}_c$,0,1;
        \EndIf
    \EndIf
	\end{algorithmic}
    \label{alg:HH-PC_IPG}
\end{algorithm}


\section{Implementation-based Analysis}

In this section, we discuss various analysis performed on our proposed solution along with the obtained results. 
The publicly available dataset MAWI20 \cite{MAWI20} is used in the performance evaluation. The Measurement and Analysis on the WIDE Internet (MAWI) traces are 15 minutes long and are collected from the backbone a Japanese academic network. Each trace has over 500 million packets. The traces are split into 1, 5, and 10 second chunks for the analysis.

Also, the proposed algorithm is evaluated on the publicly available CAIDA dataset \cite{CAIDA19url}. CAIDA's passive traces dataset contains real data traces collected from high-speed monitors on a commercial backbone link. In the CAIDA data set, specifically, the restricted access 2019 CAIDA data consists of more than 2 billion packets that span 1 hour. We used a fraction of this dataset, consisting of 29 million packets, in our analysis.

\begin{figure}[hbtp]
\centering
\includegraphics[width=0.47\textwidth]{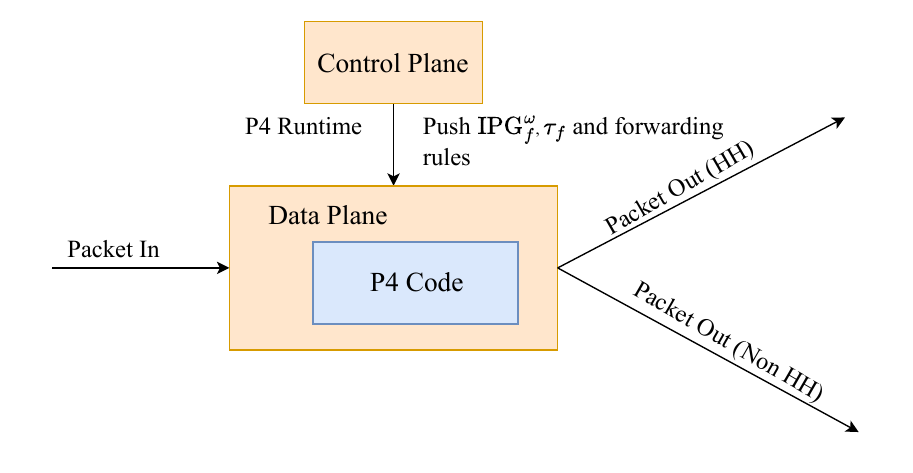}
 \caption{High level system architecture.}
    \label{fig:NA}
\centering
\end{figure}

The high level system architecture overview of placement of our algorithm is shown in Fig.~\ref{fig:NA}. Every time a packet enters the switch, the data plane where the proposed algorithm resides will make the decision whether the packet belongs to a Heavy Hitter flow or not. The necessary tables to make these decisions will be pushed by the control plane using the Equation~\ref{eq:1}. The proposed approach has been implemented in P4 code for the target Intel Tofino1 WEDGE-100B switch, setup in the International Center for Advanced Internet Research (iCAIR), Northwestern University.   
Initially, during network setup, the  control plane pushes the normal packet forwarding rules and  pushes the $\tau$ values for the corresponding $\mathrm{IPG}$ values  \cite{hh-ipg}. We compiled the proposed algorithm on Tofino using P4 Studio SDE 9.13.4. 

Using the P4 Insight tool, it was seen that the algorithm used 11 stages and 385 clock cycles, out of which 217 cycles are due to the latency introduced by the proposed algorithm. The Tofino hardware resources used by the proposed algorithm is presented in the Table~\ref{tab:hwRes}. These are the worst case figures among all 11 stages. In addition, storage for the hash table was required: this had $m$  memory slots with each slot containing a 32-bit flow ID, 16-bit weighted IPG, 16-bit last timestamp value, 8-bit $\tau$ value, (72 bits similar to \cite{hh-ipg}) and an extra Packet Count value of $l$ bits. The $l$ value depends on the threshold for Packet Count; for example, if threshold is 1024, $l$ is 10 bits. Hence total memory occupied is $m*(72+l)$ bits. The resource requirements of the proposed algorithm are thus seen to be well within the resource availability and hence we can additionally implement normal switching and custom functions based on requirement along with proposed algorithm.



\begin{table*}[tbh]
    \centering
    \caption{Tofino Resources used by the proposed algorithm}
    \label{tab:hwRes}
    \begin{tabular}{|r|r|r|}
    \hline
     \textbf{Resource} &   \textbf{Max usage (\%)} &   \textbf{Mean usage (\%)} \\\hline\hline
     Exact Match Input Crossbar &   10.2 & 3.1 \\\hline
     Hash Distribution Unit &    50.0  & 11.1\\\hline
     Exact Match Result Bus &    25.0  & 9.4\\\hline
     Action Data Bus Bytes & 6.3 & 2.3\\\hline
    \end{tabular}
\end{table*}     

For the detailed performance studies provided below, the mechanisms has been implemented using a standalone Python-based program on a machine with 12-core, 12th Gen Intel\textregistered\ Core\textsuperscript{TM} i7-12700 CPU up to 4.9~GHz and 16~GB DDR5 memory. The artifacts for both targets are made available in our Github repository \cite{CodeGit}. 

The standard machine learning (ML) evaluation metrics for heavy hitter detection algorithms were considered as in \cite{hh-ipg}, \cite{ChainSketchTON23}. These include : 

True Positive (TP) : Count of correctly classified Heavy Hitter(HH) flows. 

False Positive (FP): Count of non-Heavy Hitter flows detected as Heavy hitter. 

FN (False Negative): Count of Heavy Hitters detected as Non-HH.

Precision: Ratio of true heavy flows found over all reported flows. Precision quantifies how many of the instances the model labeled as HH were actually a HH. A high precision means that when the model predicts a flow as HH, it is likely to be a HH. 
\begin{equation}
\mathrm{Pr} = {\displaystyle\frac{{\mathrm{TP}}}{\mathrm{TP} + \mathrm{FP}}}   
\label{eq:2}
\end{equation}

Recall: Ratio of true heavy flows found over all real heavy flows. Recall indicates how many of the actual HHs the model correctly identified. A high recall means that the model is good at finding all relevant cases.
\begin{equation}
\mathrm{R} = {\displaystyle\frac{\mathrm{TP}}{\mathrm{TP} + \mathrm{FN}}}   
\label{eq:3}
\end{equation}

F1 score: Harmonic average of Precision and Recall. The F1 score balances precision and recall. A high F1 score suggests a good balance between correctly identifying HHs and not missing any. 
\begin{equation}
\mathrm{F1} = {\displaystyle\frac{{\mathrm{2} * \mathrm{R} * \mathrm{Pr}}}{R + \mathrm{Pr}}}   
\label{eq:4}
\end{equation}

A new metric called \textit{revenue} was defined as: $R = |\mathrm{TP}|-|\mathrm{FP}|-|\mathrm{FN}|$ to capture the number of false negatives. As mentioned earlier, the objective is to reduce the number of false negatives, i.e. missing detection of heavy hitter flows that can negatively impact overall network performance.


\begin{figure}[hbtp]
\centering
\subcaptionbox{10:00 trace} 
{\includegraphics[width=0.23\textwidth]{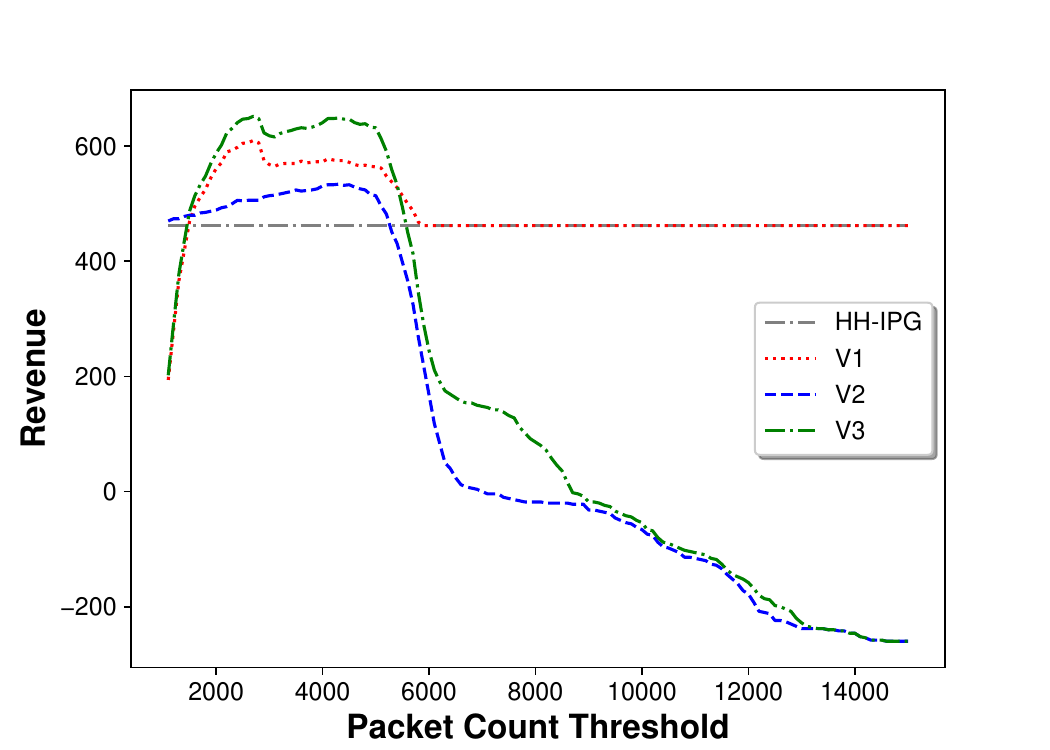}}
\hfill
\subcaptionbox{16:15 trace}
{\includegraphics[width=0.23\textwidth]{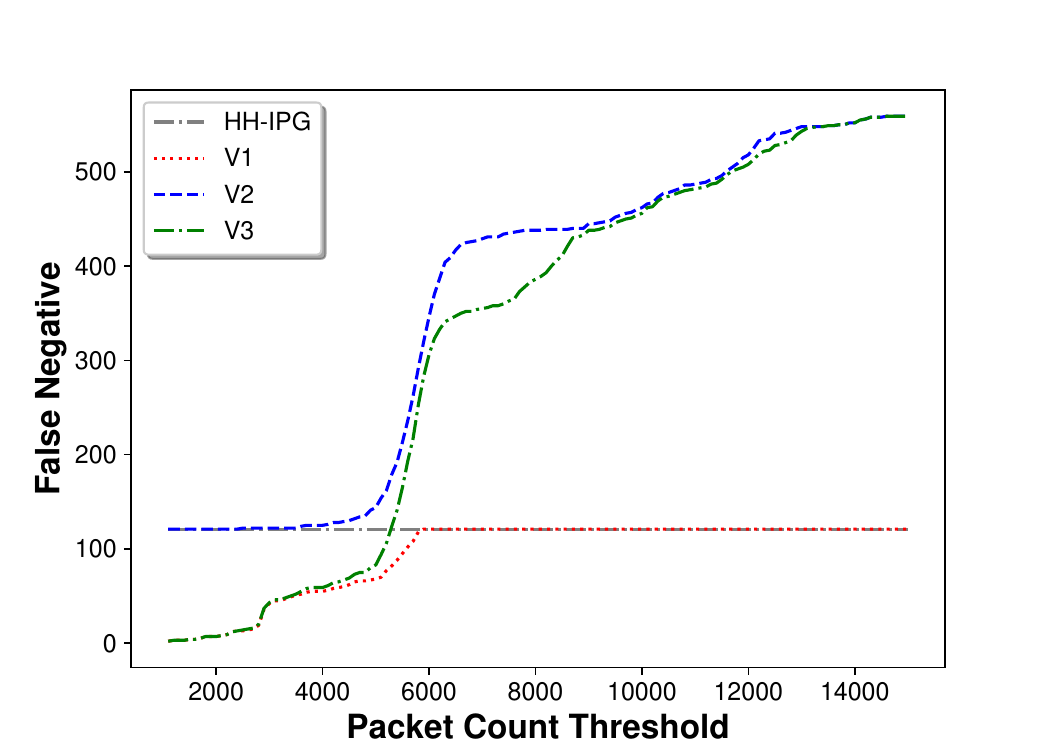}}
\caption{Varying Packet Count Threshold, for HH threshold of 10~Mbps and 5~second time window: (a) Revenue, and (b) False Negatives.}  \label{fig:PthRev}
\end{figure}

\subsection{Obtaining PC Thresholds} \label{subsec:PCth}

\begin{figure*}[hbtp]
     \centering
     \begin{subfigure}[b]{0.30\textwidth}
         \centering
         \includegraphics[width=\textwidth]{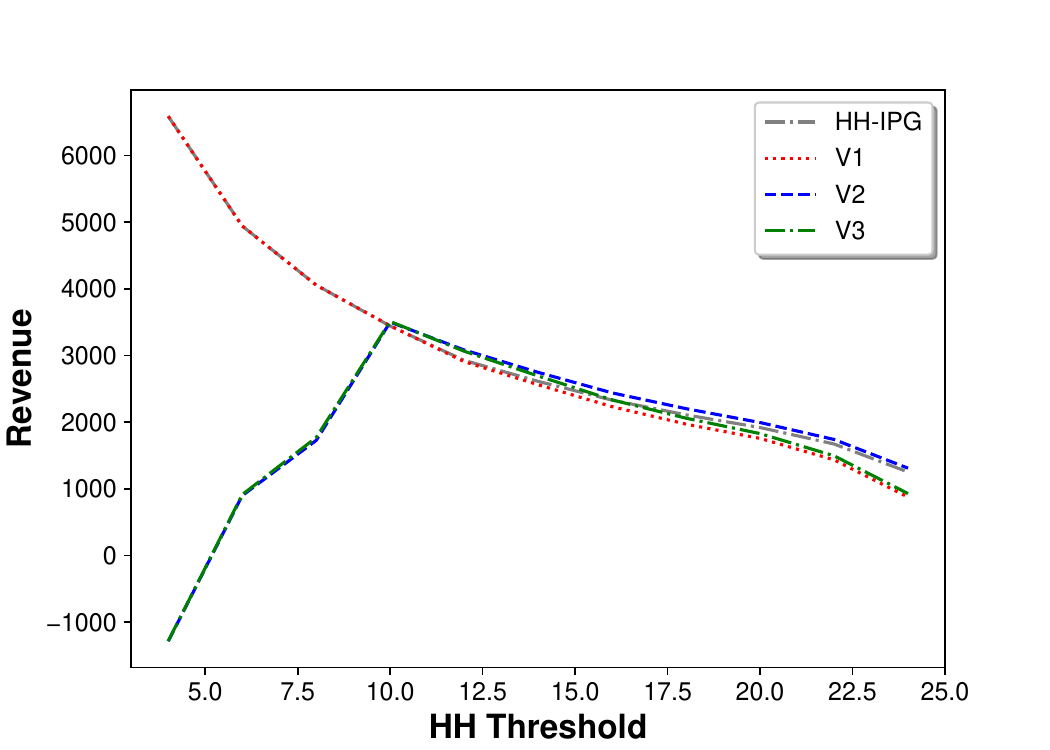}
         \caption{10:00 trace, Packet Count threshold: 1100, Time Window: 1sec}
     \end{subfigure}
     \hfill
     \begin{subfigure}[b]{0.30\textwidth}
         \centering
         \includegraphics[width=\textwidth]{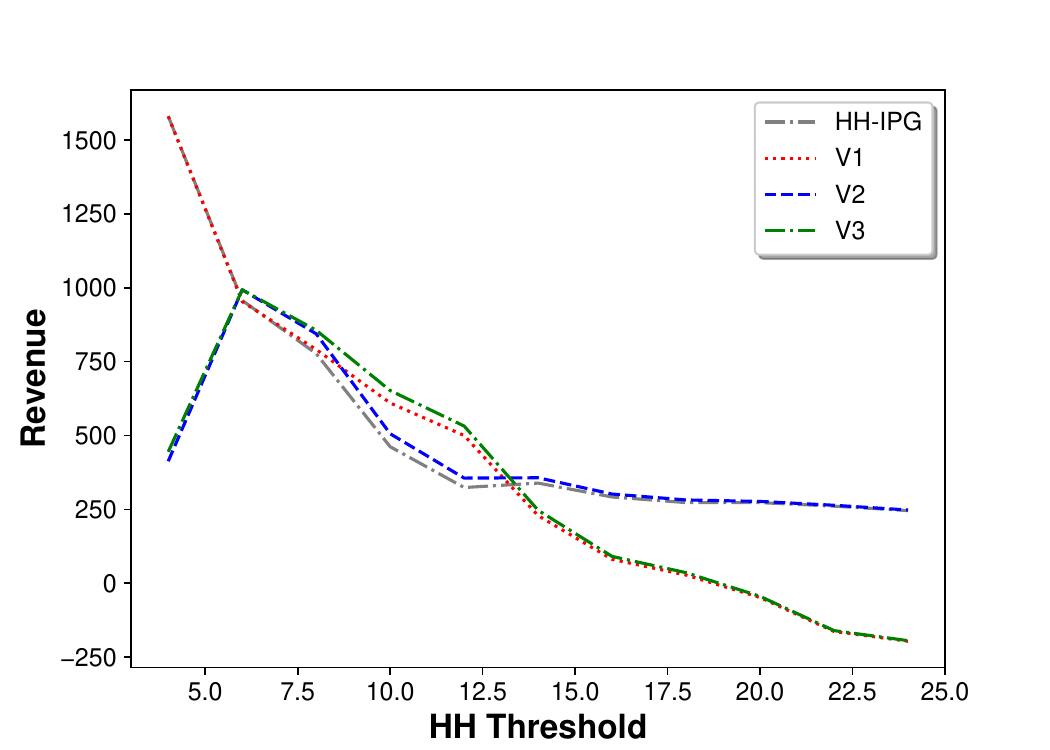}
         \caption{10:00 trace, Packet Count threshold: 2700, Time Window: 5sec}
     \end{subfigure}
     \hfill
     \begin{subfigure}[b]{0.30\textwidth}
         \centering
         \includegraphics[width=\textwidth]{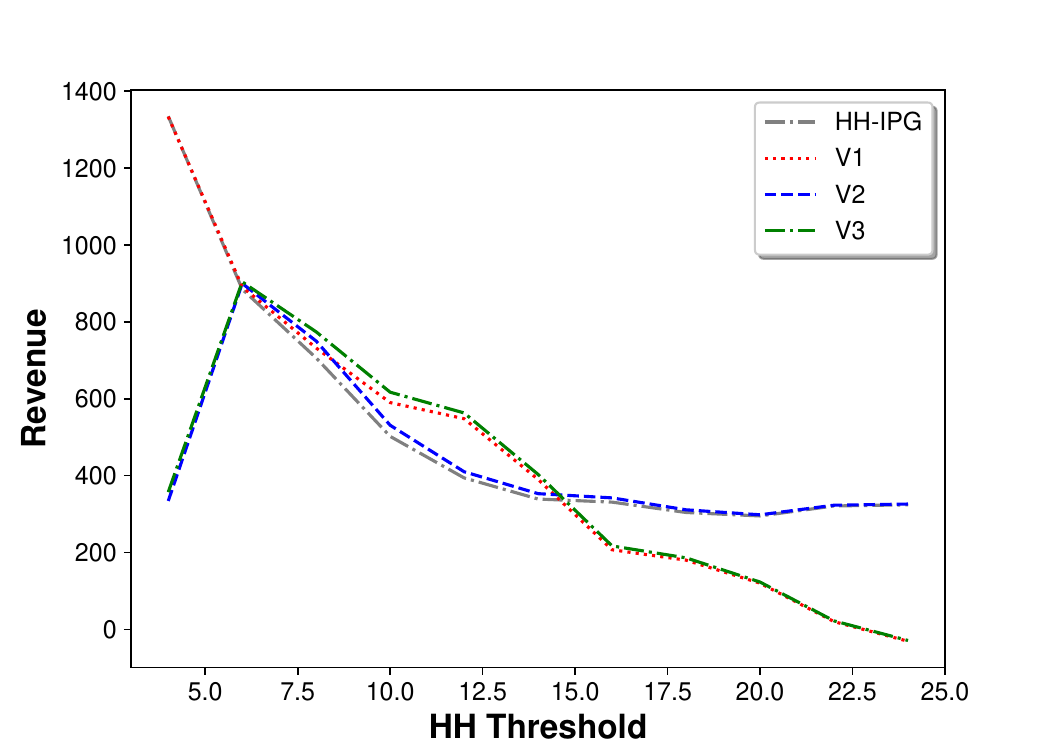}
         \caption{16:15 trace, Packet Count threshold: 2800, Time Window: 5sec}
     \end{subfigure}
        \caption{Revenue obtained, varying HH Threshold.}
        \label{fig:hhTh}
\end{figure*}


To determine suitable packet count ($\mathrm{PC}_\mathrm{th}$) threshold values, the HH flows were captured for different $\mathrm{PC}_\mathrm{th}$ values. The changes in revenue for various threshold values were measured. Further, the impact of the threshold on reducing false negatives is studied. The results for a subset of traces from MAWI20 dataset \cite{mawilab} are shown in Fig.~\ref{fig:PthRev}, where we captured HH flows for different $\mathrm{PC}_\mathrm{th}$ values in the steps of 100 between 1,100 and 20,000. Here,  $\alpha = 0.99$ as in the original HH-IPG algorithm  for small-duration time windows \cite{hh-ipg}. Packet thresholds in the range of 2000 to 6000 yielded maximum revenue, and hence we can choose any value in this range to be used as the packet count threshold.
Note that revenue does not change for original HH-IPG since it does not consider packet count. All three versions are seen to provide better performance than the IPG-based algorithm, with suitable $\mathrm{PC}_\mathrm{th}$ values chosen. The low revenue values for V2 and V3 for very high thresholds are as expected, since the PC value is used in decision making along with IPG metric. Hence, setting high $\mathrm{PC}_\mathrm{th}$ values (e.g., beyond 8000 in considered case) makes the algorithm miss heavy hitters. This increases the False Negative count, thereby reducing revenue. The V1 variant uses the PC value only during hash collision and hence performs better. 
By choosing a suitable value for $\mathrm{PC}_\mathrm{th}$, packet count till hash collision helps in improving the HH detection when used together with the IPG metric. From the experiments, it was observed that V1 is either consistently better than or as good as HH-IPG, regardless of the packet threshold values. 

\begin{figure*}[hbtp]
     \centering
     \begin{subfigure}[b]{0.30\textwidth}
         \centering
         \includegraphics[width=\textwidth]{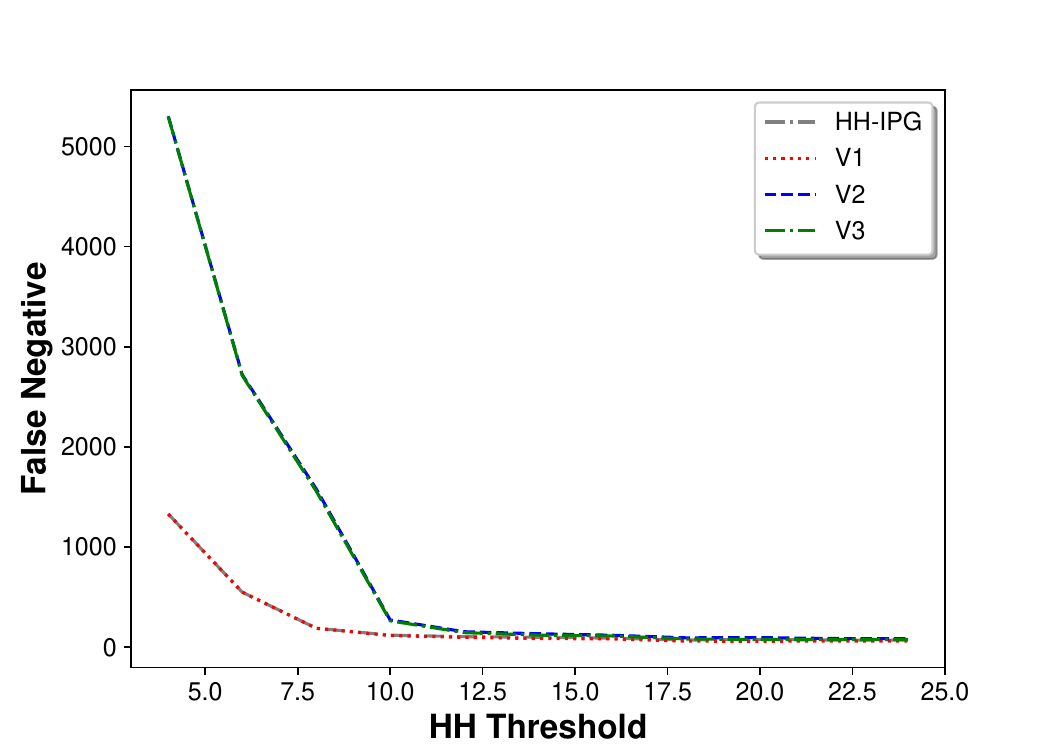}
         \caption{10:00 trace, Packet Count threshold: 1100, Time Window: 1sec}
     \end{subfigure}
     \hfill
     \begin{subfigure}[b]{0.30\textwidth}
         \centering
         \includegraphics[width=\textwidth]{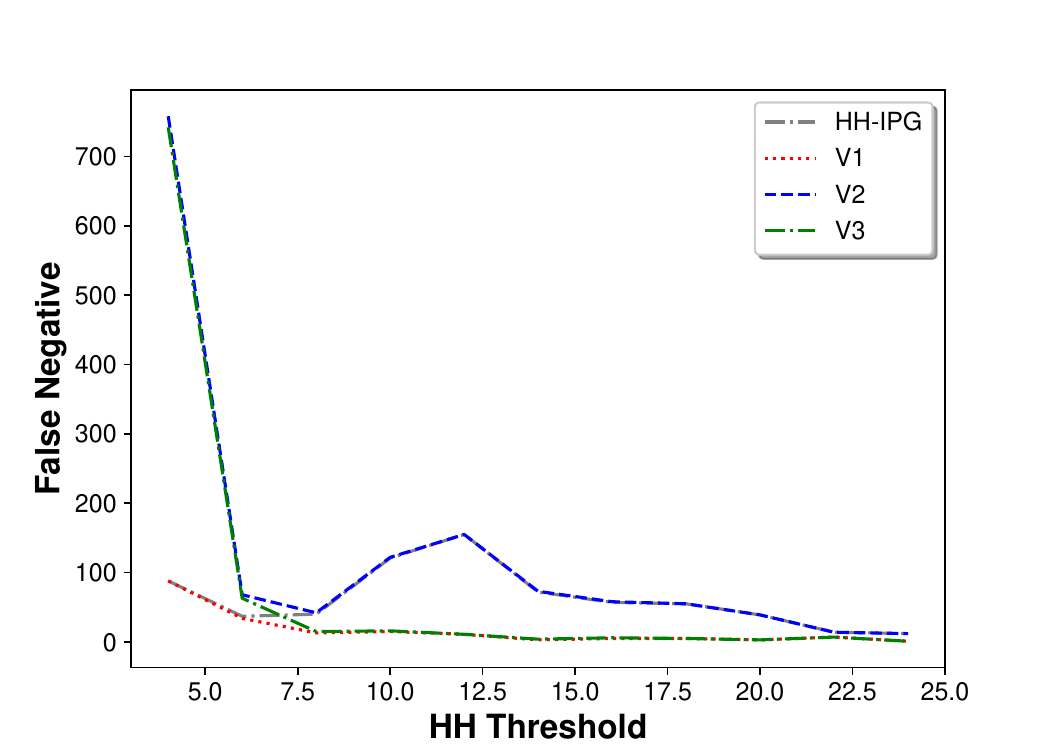}
         \caption{10:00 trace, PC threshold: 2700,  Window: 5~seconds}
     \end{subfigure}
     \hfill
     \begin{subfigure}[b]{0.30\textwidth}
         \centering
         \includegraphics[width=\textwidth]{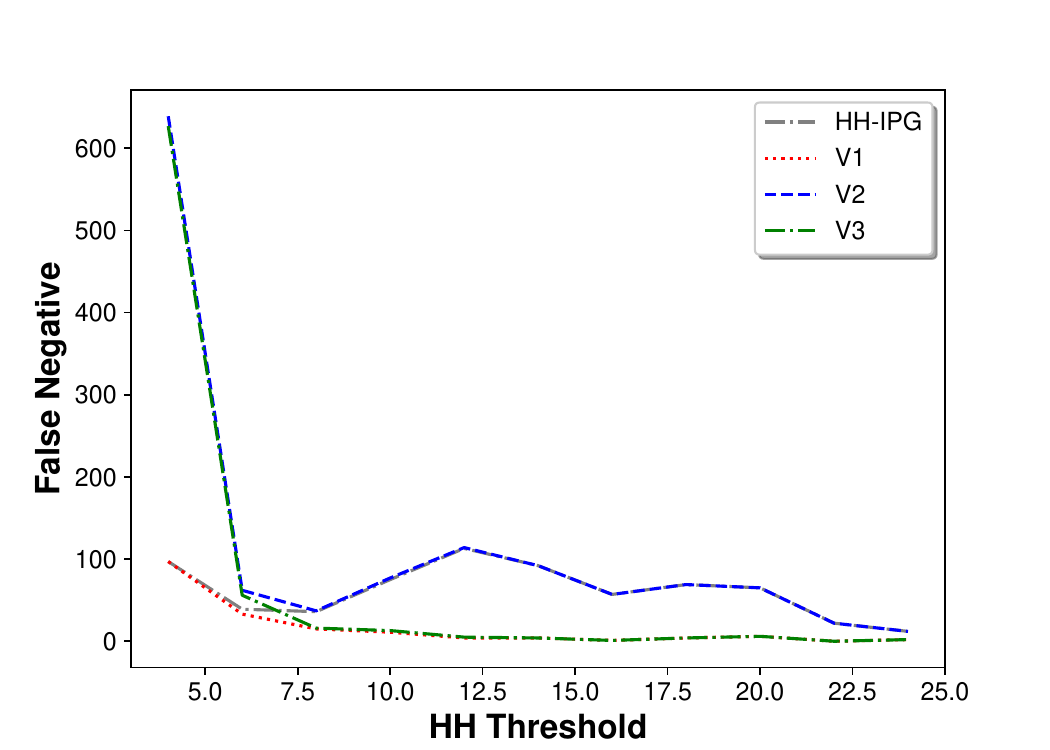}
         \caption{16:15 trace, PC threshold: 2800,  Window: 5~seconds.}
     \end{subfigure}
        \caption{Missed HHs observed, varying HH Threshold.}
        \label{fig:hhThFN}
\end{figure*}

\subsection{Varying HH threshold}

Next, the impact of heavy hitter threshold ($\mathrm{HH}_{th}$) was studied. Fig.~\ref{fig:hhTh}  and Fig.~\ref{fig:hhThFN} present the revenue and false negatives results for varying HH threshold values respectively. We chose the best ($\mathrm{PC}_\mathrm{th}$) from the experimentation results of Section \ref{subsec:PCth}. Here, it is observed that V1 performs better than HH-IPG when for HH thresholds are in the range of 7-13~Mbps. Also, the packet count threshold value is set as per the observations done from experiments conducted setting HH threshold as 10~Mbps. The packet count threshold is expected to be changed with the change in the heavy hitter cutoff. Here, the same packet count threshold is still giving the better results in considerable range of 7-13~Mbps, despite the packet count threshold value being calculated for 10~Mbps as heavy hitter threshold. This shows the flexibility of the proposed algorithm. We can also observe that the objective of reducing missed HHs has been achieved by V1 in all the cases.

\subsection{ML metrics}

\begin{figure*}[hbtp]
     \centering
     \begin{subfigure}[b]{0.30\textwidth}
         \centering
         \includegraphics[width=\textwidth]{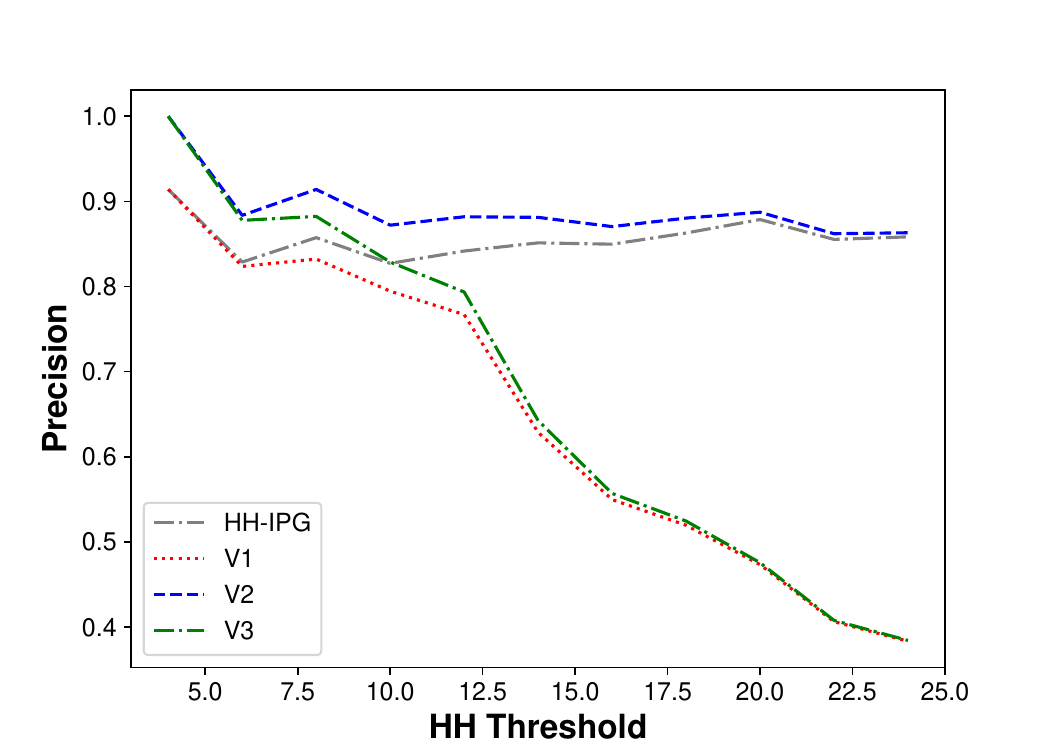}
         \caption{Precision}
     \end{subfigure}
     \hfill
     \begin{subfigure}[b]{0.30\textwidth}
         \centering
         \includegraphics[width=\textwidth]{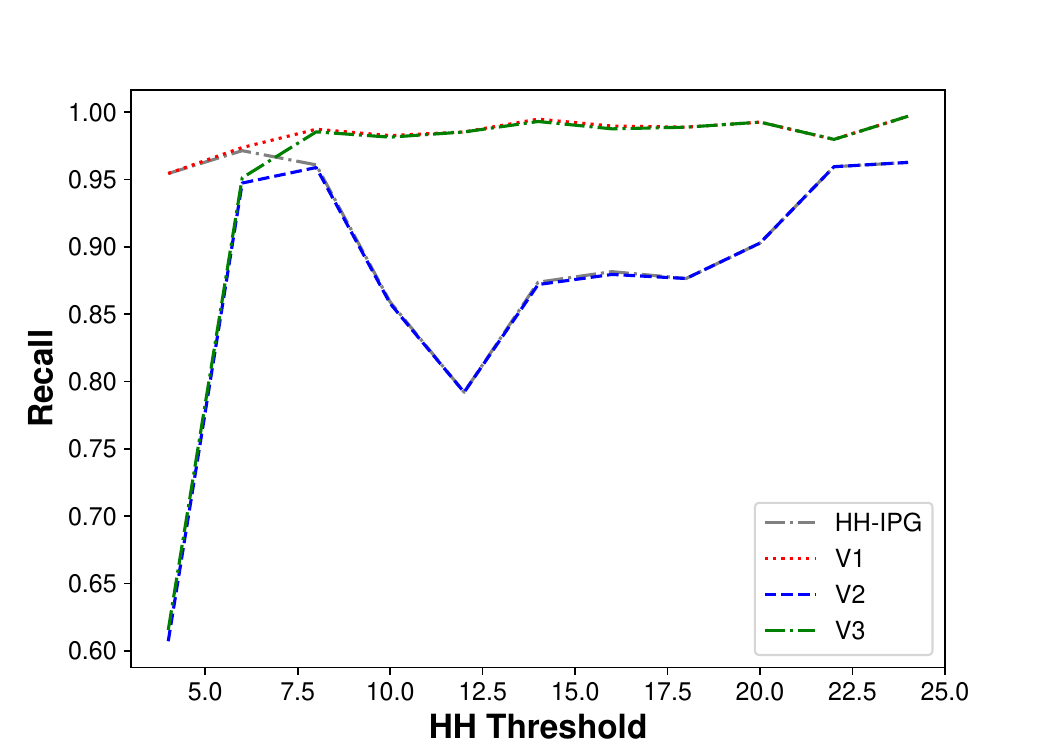}
         \caption{Recall}
     \end{subfigure}
     \hfill
     \begin{subfigure}[b]{0.30\textwidth}
         \centering
         \includegraphics[width=\textwidth]{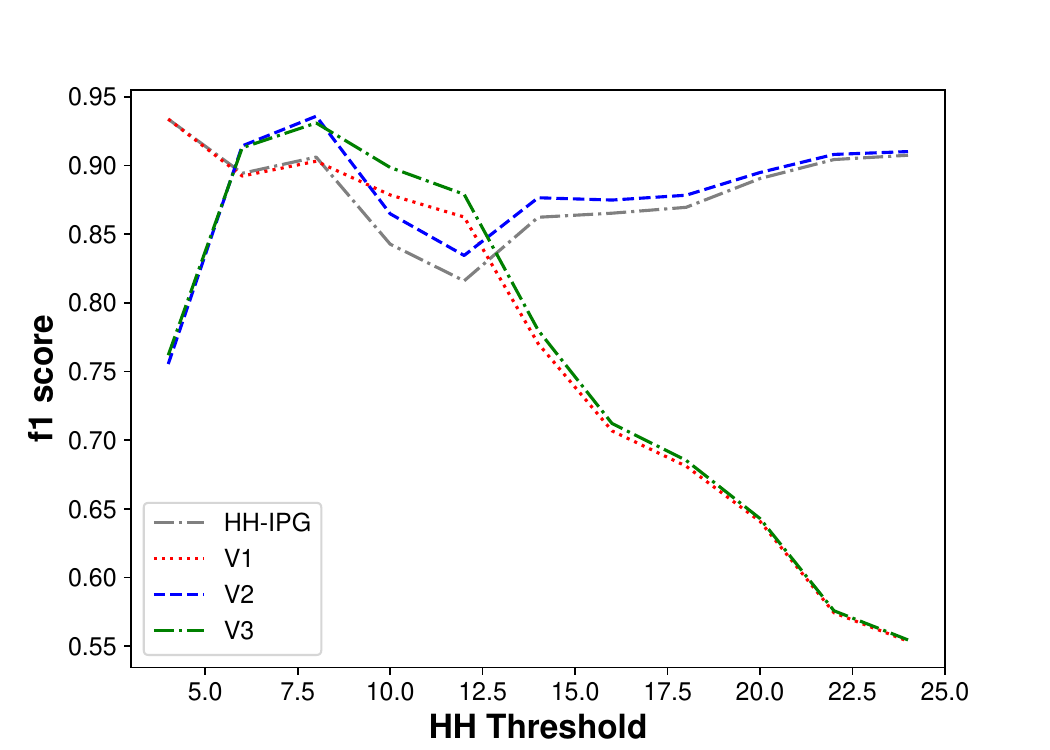}
         \caption{F1 score}
     \end{subfigure}
        \caption{Comparison of various ML metrics for 10:00 traces.}
        \label{fig:MLmetrics11}
\end{figure*}

\begin{figure*}[hbtp]
     \centering
     \begin{subfigure}[b]{0.30\textwidth}
         \centering
         \includegraphics[width=\textwidth]{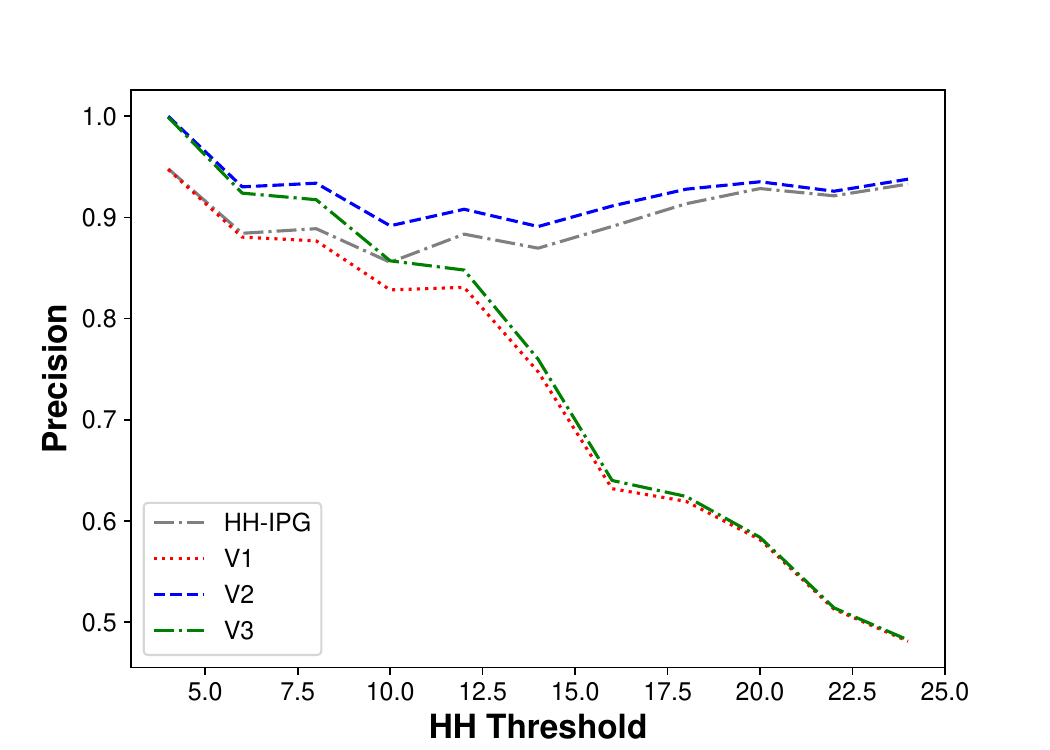}
         \caption{Precision}
     \end{subfigure}
     \hfill
     \begin{subfigure}[b]{0.30\textwidth}
         \centering
         \includegraphics[width=\textwidth]{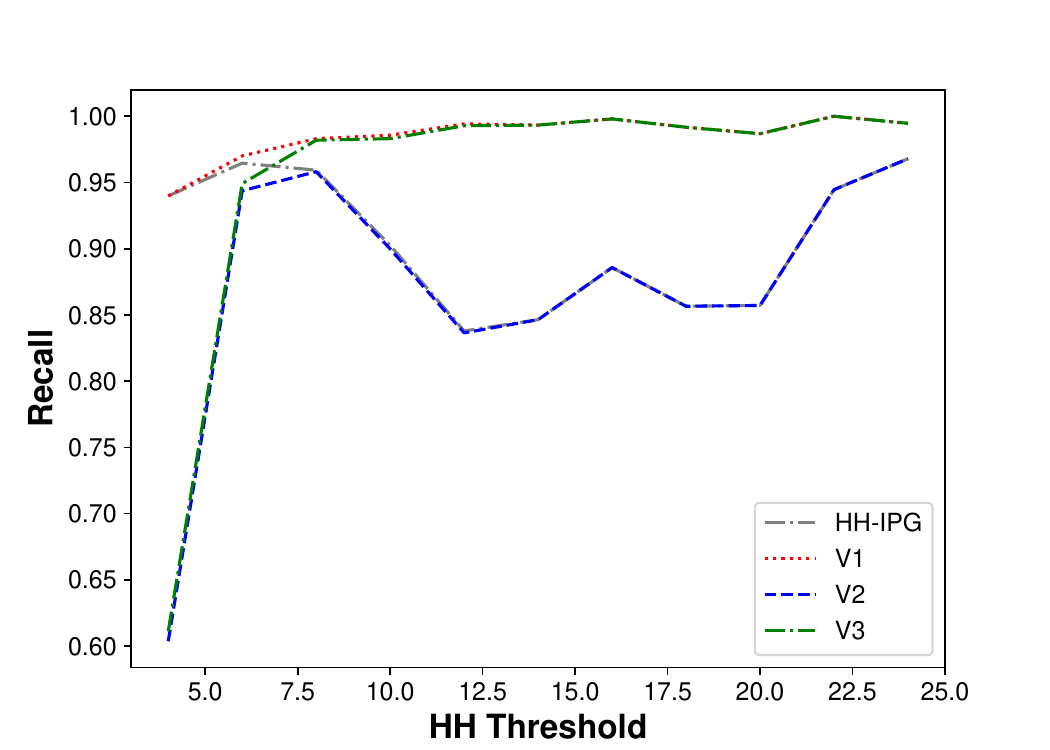}
         \caption{Recall}
     \end{subfigure}
     \hfill
     \begin{subfigure}[b]{0.30\textwidth}
         \centering
         \includegraphics[width=\textwidth]{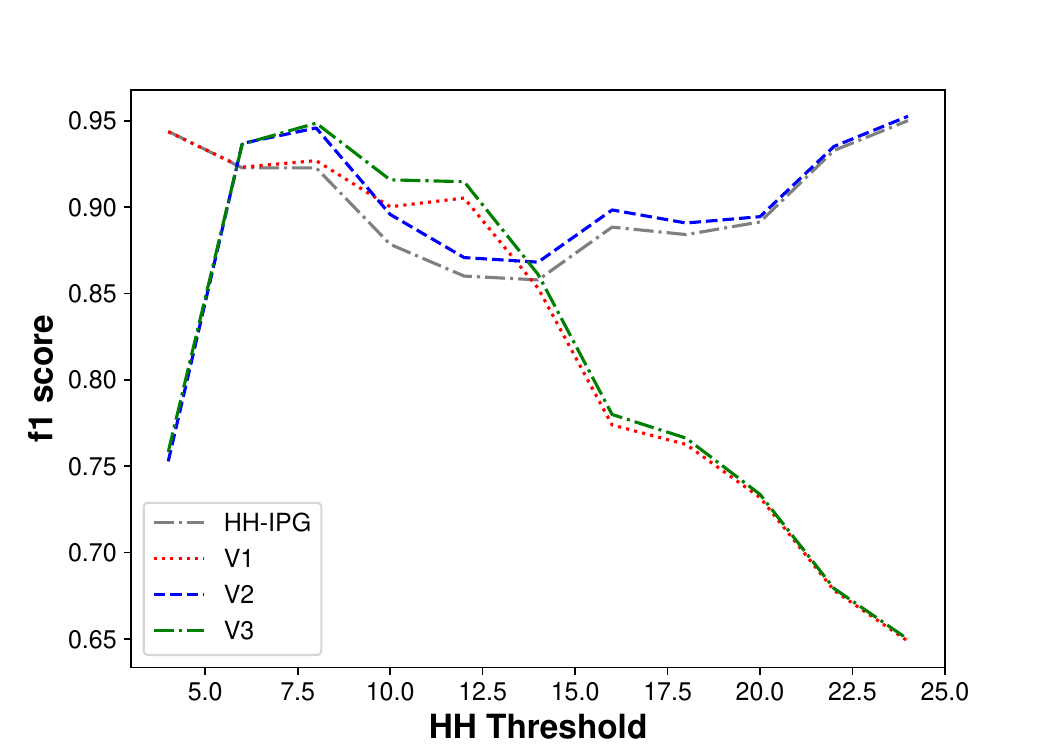}
         \caption{F1 score}
     \end{subfigure}
        \caption{Comparison of various ML metrics for 16:15 traces.}
        \label{fig:MLmetrics22}
\end{figure*}

In Fig.~\ref{fig:MLmetrics11}, the comparison of ML metrics is presented on MAWI dataset, with each 15-minute long trace divided into 5~sec time window traces. We fixed the packet count threshold as 2700. Similar to earlier discussion, the proposed algorithm versions are performing better than the baseline approach when the $\mathrm{HH}_{th}$ is 10~Mbps which is the originally considered threshold while finding best $\mathrm{PC}_{th}$. Along with that, we can also observe here that V1 performs better for the HH thresholds ranging from 5 to 13~Mbps in terms of all the three ML metrics. Hence, the proposed algorithm can give better results despite some errors with respect to the packet count threshold value. We can observe here that V1 performs better for the HH thresholds ranging from 5 to 13~Mbps. For higher HH thresholds, V1 does not show any significant improvement whereas V2 shows better results. In Fig.~\ref{fig:MLmetrics22}, ML metrics are compared for 16:15 trace divided into 5~sec time window traces with fixed packet count threshold 2800. The trends are seen to be similar.

\begin{figure}[hbtp]
\centering
\includegraphics[width=0.49\textwidth]{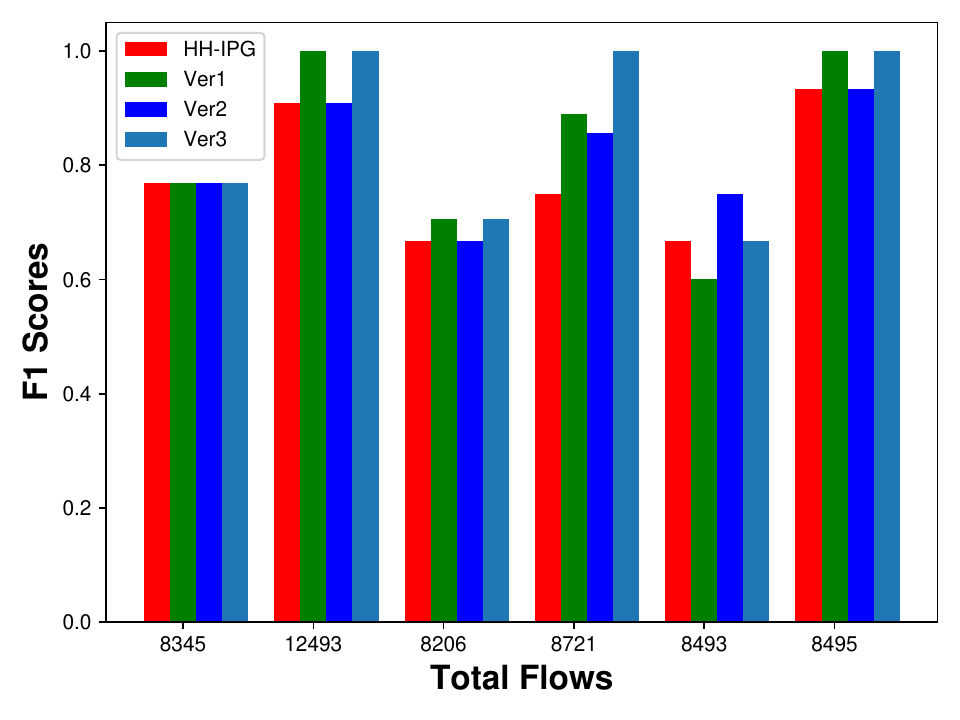}
 \caption{F1 score for  10:00 trace with 5~sec  window.}
    \label{fig:f1Comp}
\centering
\end{figure}

The comparison of F1 score for initial 5~sec traces with the specific case of $\mathrm{PC}_{th}$ as 2700 and  $\mathrm{HH}_{th}$ as 10~Mbps is presented  in Fig.~\ref{fig:f1Comp} and in Table~\ref{tab:f1}. For the overall 15-minute trace, the F1 score of the proposed scheme is reasonably higher compared with the HH-IPG algorithm. Quantitatively, the F1 score for the baseline approach was 0.84276 for the considered trace, which increased to 0.875843 in version 1, to 0.881888 in version 2, and to 0.912665 in version 3 of the proposed approach with best $\mathrm{PC}_{th}$. It is also observed that the F1 score is not always higher for all the small time windows. It is seen that in some of the small time windows, only the IPG metric gives better results. However, the overall aggregated results across the larger trace duration show that the PC feature improves the F1 score.


\begin{table*}[tbh]
\centering
\caption{F1 score comparison for  5~second window traces}
\label{tab:f1}
\begin{tabular}{|p{1.2cm}|p{1.5cm}|p{1.5cm}|p{1.5cm}|p{1.5cm}|p{1.5cm}|}
\hline
 Win. \# &   \textbf{Total Flows} & \textbf{HH-IPG} & \textbf{V1} & \textbf{V2} & \textbf{V3}\\\hline\hline
 1 &   8,345 & 0.7692 & 0.7692 & 0.7692 & 0.7692 \\\hline
 2 &    12,493 & 0.9090 & 1.0000 & 0.9090 & 1.0000 \\\hline
 3 &    8,206 & 0.6666 & 0.7059 & 0.6667 & 0.7059 \\\hline
 4 &    8,721 & 0.7500 & 0.8889 & 0.8571 & 1.0000 \\\hline
  5 &   8,493 & 0.6667 & 0.6000 & 0.7500 & 0.6667 \\\hline
  6 &   8,495 & 0.9333 & 1.0000 & 0.9333 & 1.0000 \\\hline
  7 &   8,083 & 1.0000 & 0.9090 & 1.0000 & 0.9090\\\hline
  8 &   8,166 & 0.7500 & 0.6667 & 0.7500 & 0.6667\\\hline
 9 &    8,604 & 0.8333 & 1.0000 & 0.7272 & 0.9230 \\\hline
  10 &   8,395 & 0.8889 & 0.8889 & 0.8889 & 0.8889 \\\hline
 11 &    8,387 & 0.8000 & 0.7272 & 0.8889 & 0.8000 \\\hline
  12 &   8,673 & 0.7500 & 0.7500 & 0.8571 & 0.7500 \\\hline
 13 &    8,166 & 0.8889 & 0.8889 & 0.8889 & 0.8889 \\\hline
 14 &    8,955 & 0.6667 & 0.7272 & 0.7500 & 0.8000\\\hline
 15 &    8,201 & 0.8000 & 0.9230 & 0.8000 & 0.9230 \\\hline
\end{tabular}
\end{table*}

\subsection{Optimizing the results based on requirements} \label{Sec:OptReq}
The heavy hitter detection algorithm has versatile applications (Section~\ref{sec:Intro}). Hence, based on application, we may need to optimize our results for a particular evaluation metric. As an example, consider a machine learning tool which is flagging the possible threats which later will be investigated manually by a threat hunter. Here, false positives are acceptable to certain extent as it will be discarded by threat hunter later; but false negatives are not acceptable. Similarly, there might be some application that requires optimization with respect to the F1 score metric.
Taking this into account, we are testing whether our algorithm can accommodate such requests by optimizing our algorithm with respect to a different metric, the F1 score, in this section. To do so, we plot the F1 score values obtained by our algorithm for various packet count thresholds.

\begin{figure}[hbtp]
\centering
\subcaptionbox{wrt. Revenue} 
{\includegraphics[width=0.23\textwidth]{FinalNimbusSans/10/Ft-resultsResV1-3_hhTh10Win5.txtRevenue_2.pdf}}
\hfill
\subcaptionbox{wrt. F1 score}
{\includegraphics[width=0.23\textwidth]{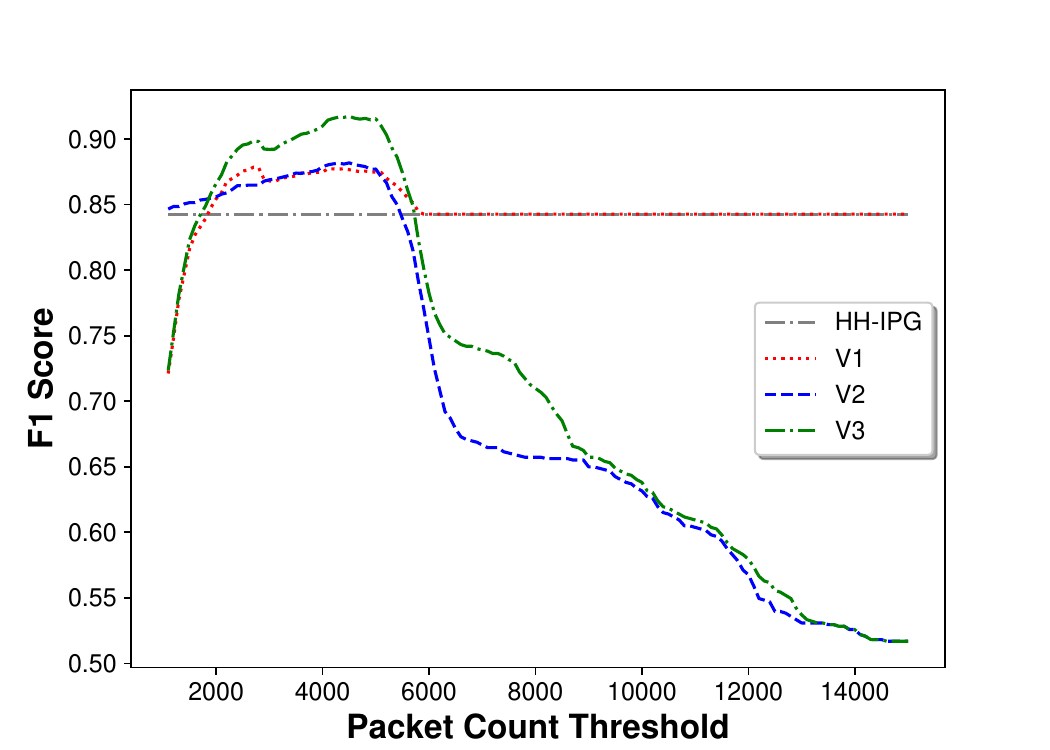}}
\caption{Finding the optimal packet count threshold value (a) wrt. Revenue, and (b) wrt. F1 score}  \label{fig:optif1}
\end{figure}

Comparing the above results with the earlier results where optimization is done with respect to revenue metric, we can observe that the algorithm behaves similarly as shown in Fig.~\ref{fig:optif1}. The pattern is similar because the same dataset is used for experimentation, which results in the exact same value of true heavy hitters and the total number of flows.

\subsection{Comparison of results for various hash functions} \label{Sec:diffHash}
To understand the impact of the hash function being used on the results of our algorithm, we observe the HH detection by our algorithm while using 3 different hash functions. Considering the computational limitations of PDP switches, we limit our study to the linear functions. 

\begin{figure}[hbtp]
\centering
\subcaptionbox{Hash 1} 
{\includegraphics[width=0.23\textwidth]{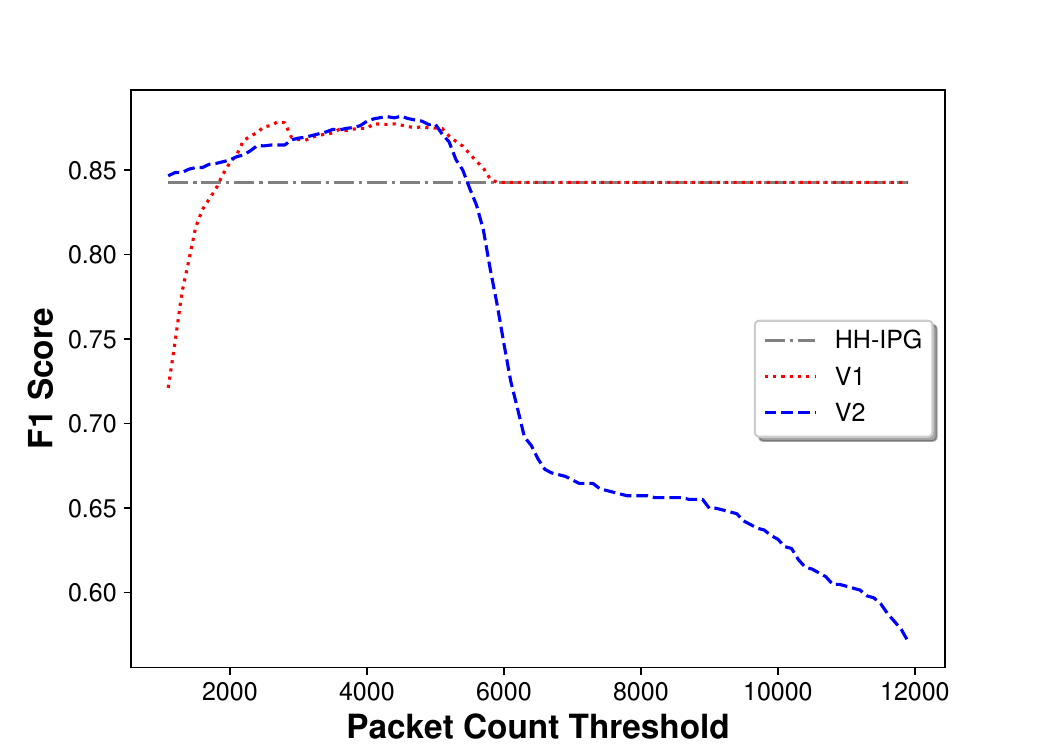}}
\hfill
\subcaptionbox{Hash 3}
{\includegraphics[width=0.23\textwidth]{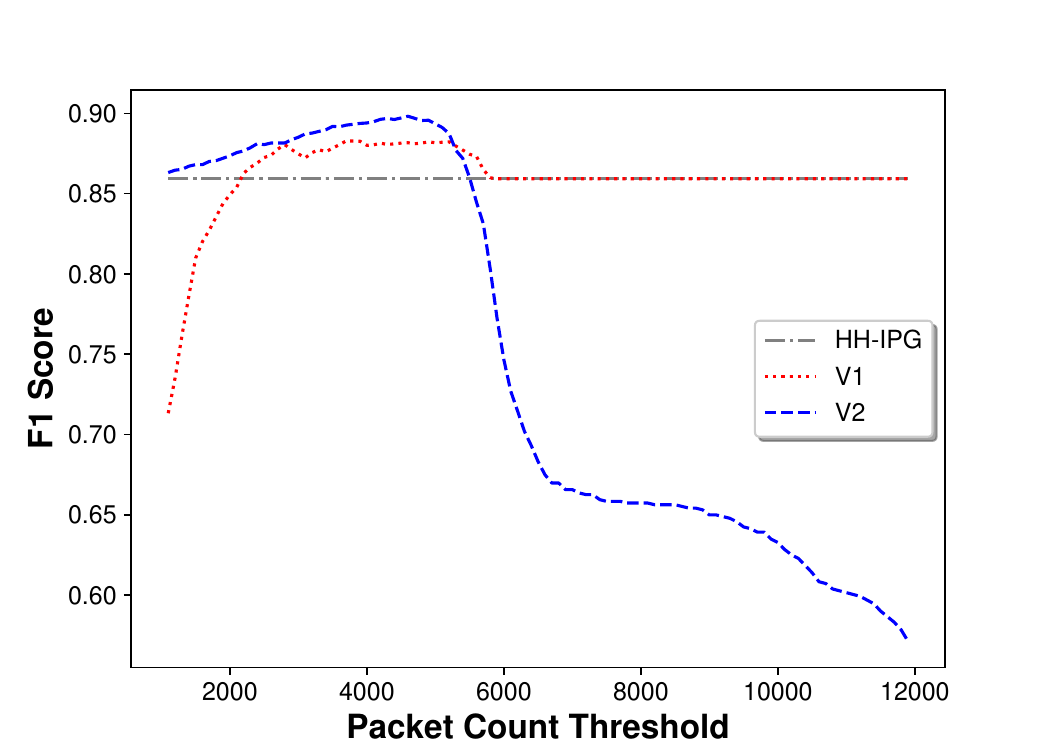}}
\caption{Comparison of how each algorithm behaves wrt. chosen Hash functions (a) Hash 1 and (b) Hash 3}  \label{fig:hashCmp}
\end{figure}

\begin{figure}[hbtp]
\centering
\subcaptionbox{Version 1} 
{\includegraphics[width=0.23\textwidth]{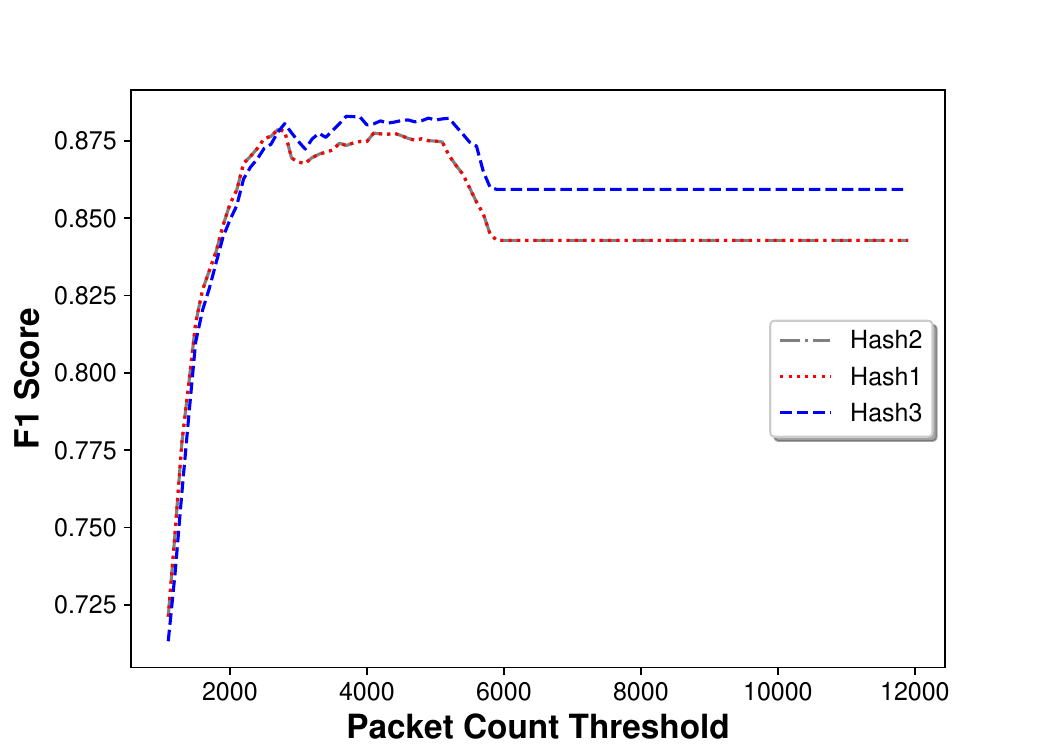}}
\hfill
\subcaptionbox{Version 2}
{\includegraphics[width=0.23\textwidth]{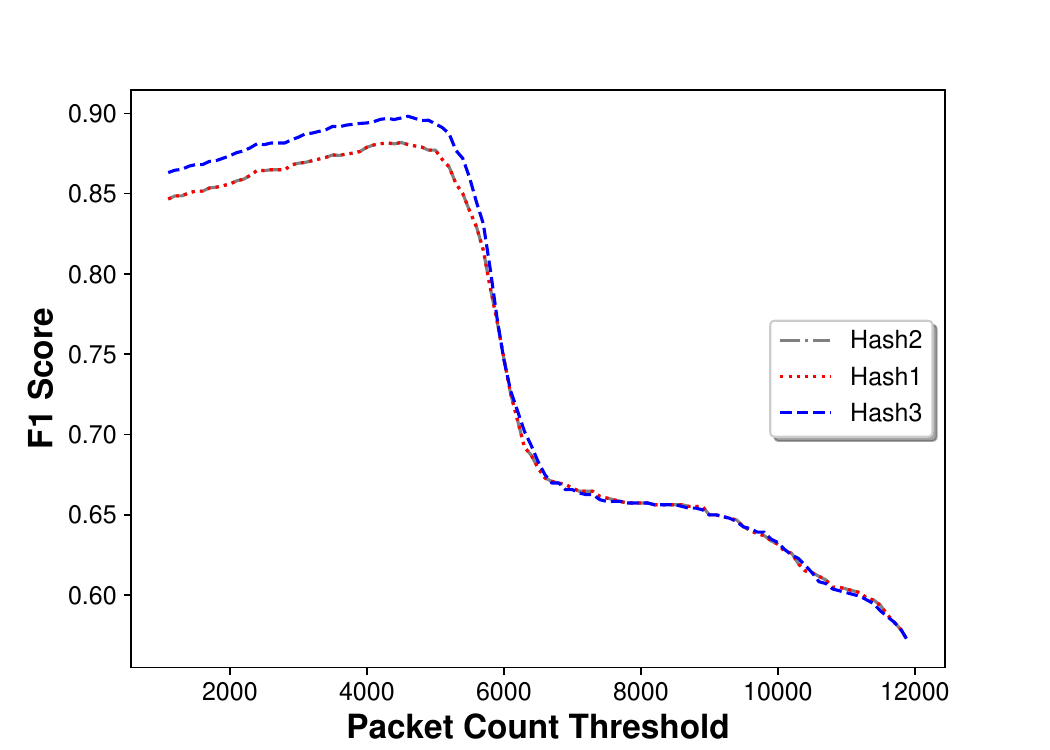}}
\caption{Comparison of how selected algorithm (i.e (a) Version 1 and (b) Version 2) behaves wrt. different Hash functions.}  \label{fig:hash123}
\end{figure}

We used 3 different hash functions: (i) Modulo Hashing or Division Method (Hash 1); (ii) Multiplicative Hashing (Hash 3); (iii) Direct Mapping (Hash 2). The results for how each version behaves for these different linear hash functions are presented in Fig.~\ref{fig:hashCmp}. Here, we can observe that the algorithm gives similar results despite the hash function being used and also the packet count threshold value for which the algorithm gives the best result also follows a similar pattern giving the better results around the range 2200 to 5500.
Also, regardless of the hash function being used, each algorithm behaves almost similarly. This is observed from the experimentation results shown in Fig.~\ref{fig:hash123}, where, for the chosen hash function, a comparison of how the F1 score varies with different algorithms is presented.

\subsection{Comparison of results for various data structure size} \label{Sec:DSsize}

For any algorithm running in a PDP switch, the memory utilization by the algorithm is very important, as the dynamic memory available for the custom algorithm will be limited in many PDP switches. Hence, we study the behavior of the algorithm for various allocated memory sizes for the custom data structure being used in our algorithm. We considered data structure sizes to be 500, 1000, 1500, 2000, 2500, 3000, and 5000 entries.

\begin{figure}[hbtp]
\centering
\subcaptionbox{Version 1} 
{\includegraphics[width=0.23\textwidth]{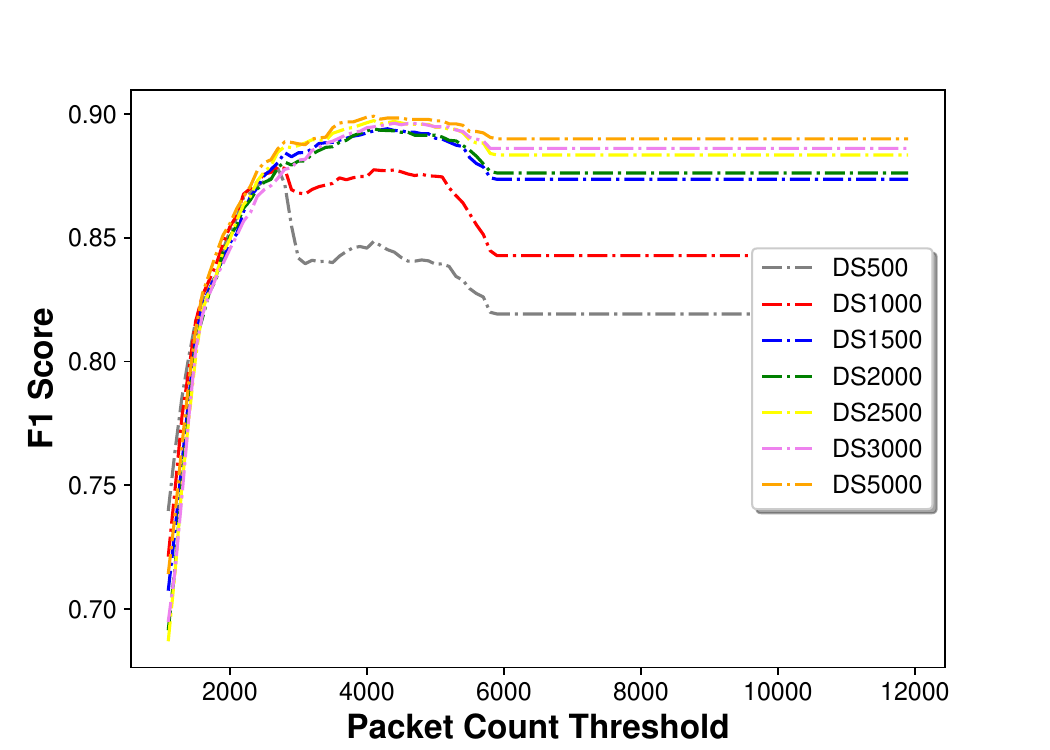}}
\hfill
\subcaptionbox{Version 2}
{\includegraphics[width=0.23\textwidth]{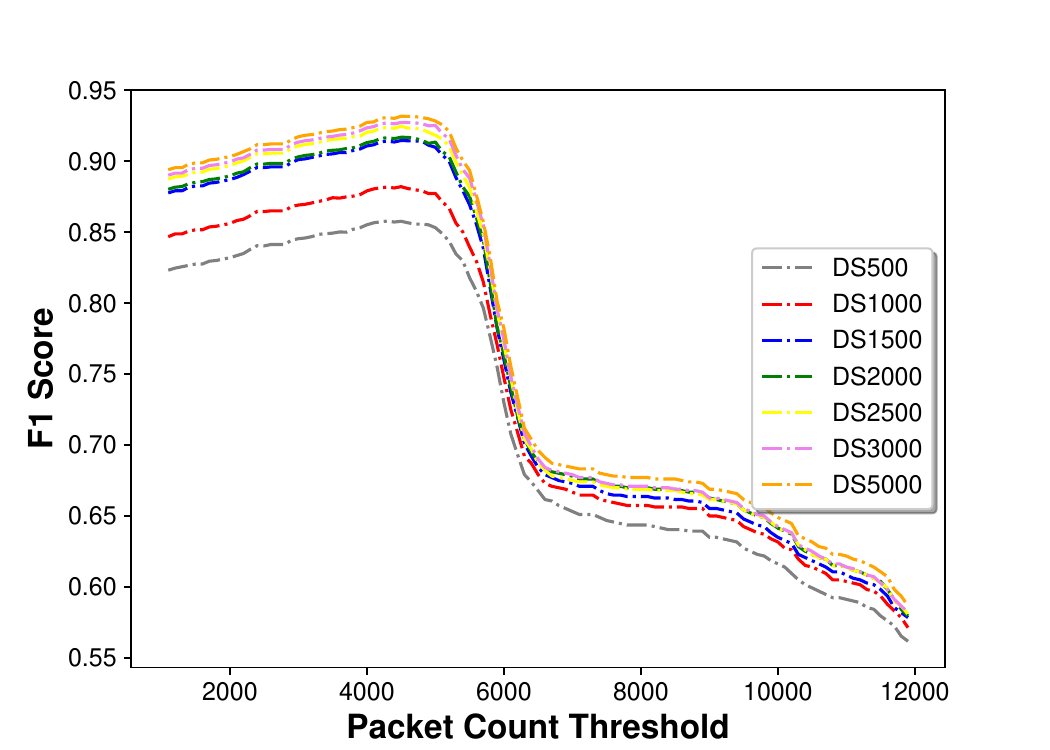}}
\caption{Comparison of how each algorithm behaves wrt. different Table Sizes}  \label{fig:DSsizeCmp}
\end{figure}

The analysis results are shown in Fig.~\ref{fig:DSsizeCmp}. Here, we can observe that with an increase in the allotted size of the data structure, the F1 score increases. The results follow the expectation that given the more memory storage for data structure, details of a relatively larger number of flows can be saved in the data structure. Thus, there will be less hash collisions, and the F1 score will be better. More heavy hitters can be captured at a given time instance, and hence the results will be better.

Also, from Fig.~\ref{fig:DSsizeCmp}, we can observe that for our algorithm as well as for HH-IPG, after reaching the F1 score around 0.9; the gain by having higher memory space for the data structure will relatively decrease as we increase the table size and hence conveys that we can reasonably use table size around 3000.

\begin{figure}[hbtp]
\centering
\subcaptionbox{Hash Size 3000 entries} 
{\includegraphics[width=0.23\textwidth]{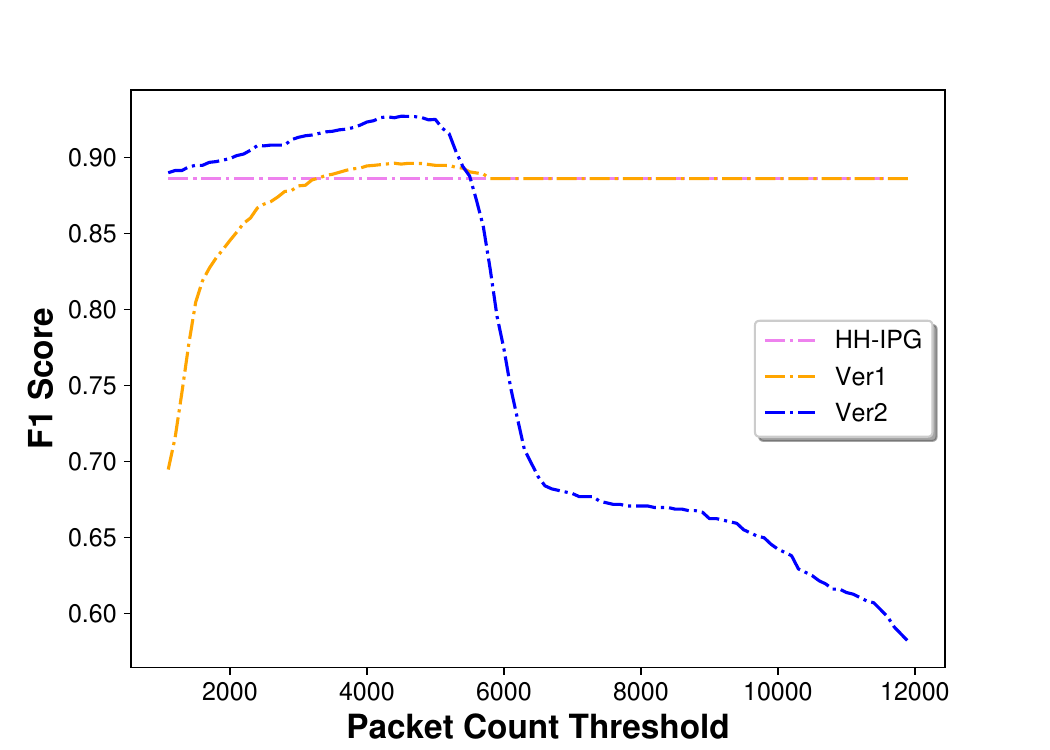}}
\hfill
\subcaptionbox{Hash Size 5000 entries}
{\includegraphics[width=0.23\textwidth]{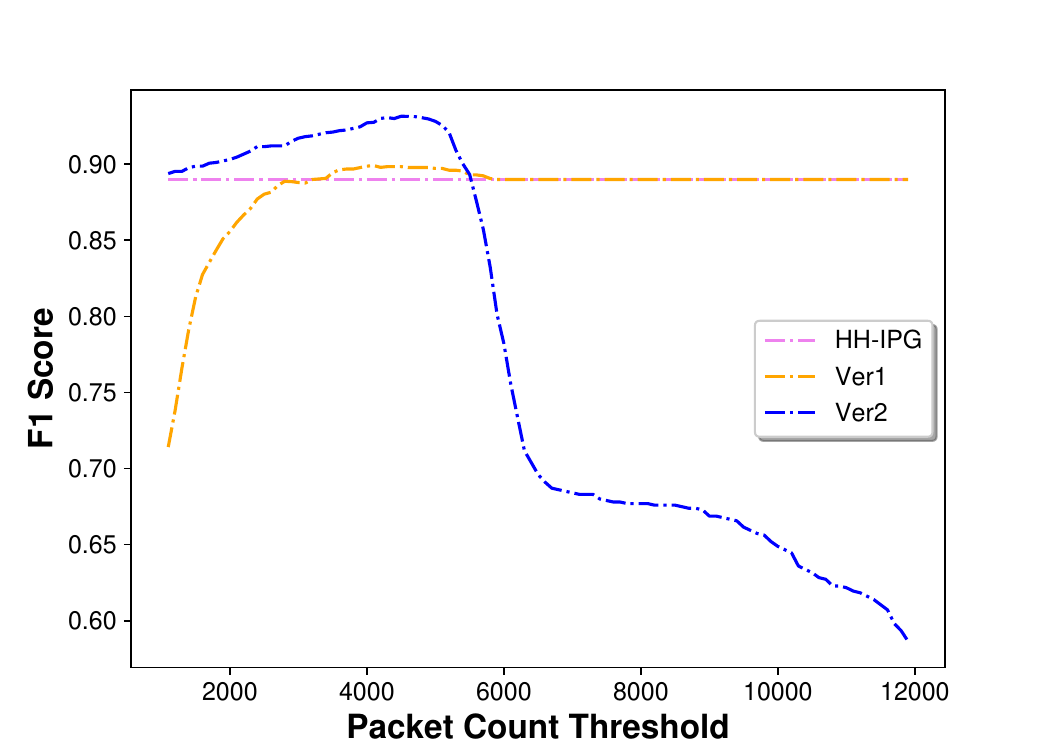}}
\caption{Comparison of F1 score for the table size of (a) 3000 (b) 5000}  \label{fig:DSsize-VerCmp}
\end{figure}

In Fig.~\ref{fig:DSsize-VerCmp}, comparison of behavior of different algorithms for different table sizes has been provided. Here, we can observe that despite the increase in the F1 score, the pattern remains similar irrespective of the table size chosen; for properly chosen thresholds, the proposed algorithms give better results than the HH-IPG. 

\subsection{Double hashing} \label{Sec:doubleHash}
As another common solution for the hash collision is using double hash, we made use of two hash functions to see whether the double hash approach could give better results. The approach used for this is that the modulo hashing (Hash 1) is applied first on the 5-tuple to get the hash table index. Here, if an entry already exists in the table with a different flow ID in the hashed entry; the second hash function 'Multiplicative hashing' is used.   

In our approach, there will be no deletion of the entry in the hash table, avoiding the deletion problem that occurs when using double hashing. Any existing entry can just be replaced by a new entry in case of hash collision at both the hash functions. The replacement is done in place of the entry pointed by the second hash function. 

\begin{figure}[hbtp]
\centering
\includegraphics[width=0.45\textwidth]{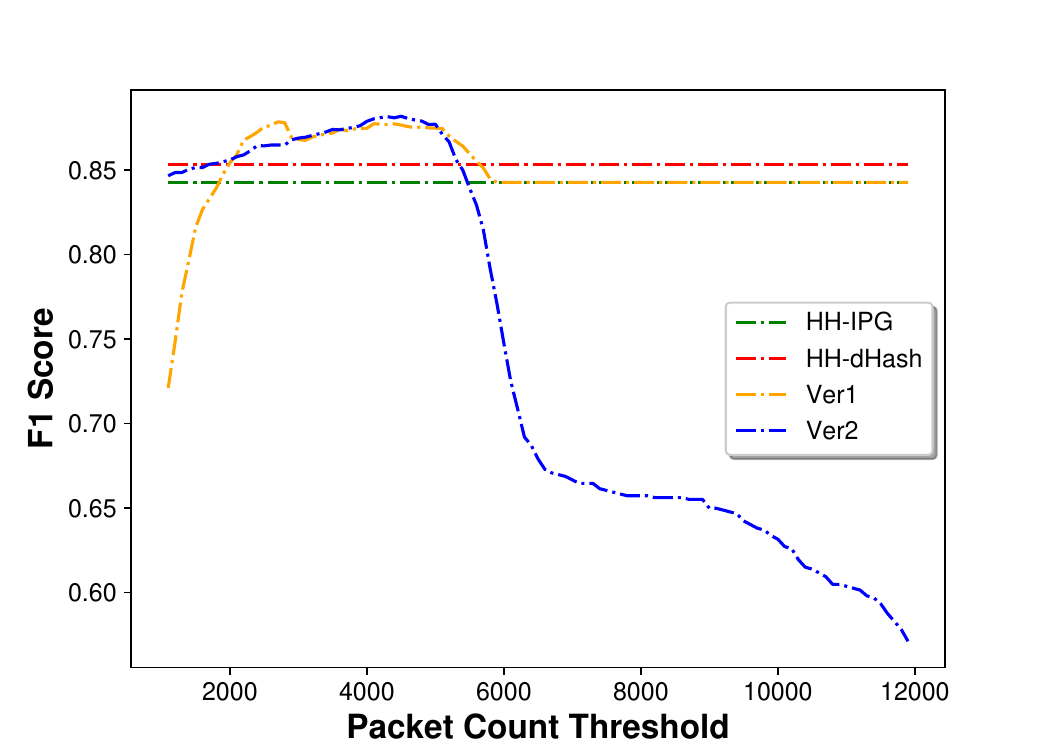}
 \caption{Comparison of proposed approach against using double hash}
    \label{fig:f1CompDhash}
\centering
\end{figure}

The comparison of result of using double hashing during hash collision is provided in Fig.~\ref{fig:f1CompDhash}. The HH-dHash in Fig.~\ref{fig:f1CompDhash} corresponds to the result of double hashing, which improved F1 score of HH-IPG algorithm from 0.842767 to 0.85322. The proposed algorithm can perform better than using double hashing. The proposed algorithm can achieve F1 score above 0.9, with the proper choice of packet count threshold. 

The above mentioned analysis is done by simulation. But we also need to consider that in case of implementation of double hashing on real programmable switch, there will be limitations on using the hash table making adverse effects on the switch functioning. For example, in case of Tofino switch, to get into a different entry in the hash table after applying second hash function; the packet needs to be recirculated. This is because of the architectural limitation of the Tofino switch. When a packet enters the Tofino switch, only one table entry can be fetched per packet in the packet's lifetime. Hence, recirculation is necessary, which will increase the traffic on the switch. The increase in traffic hinders the standard functionality of the switch, causing a decrease in performance compared with the above-shown simulation results. In simulation, we directly fetch the second entry in case of hash collision. Note that the accuracy and F1 measurement metrics will remain the same, but the time of processing, traffic congestion, and so on will be affected. 

\subsection{Results on CAIDA dataset} \label{Sec:moreDataset}

\begin{figure}[hbtp]
\centering
\subcaptionbox{F1 Score} 
{\includegraphics[width=0.23\textwidth]{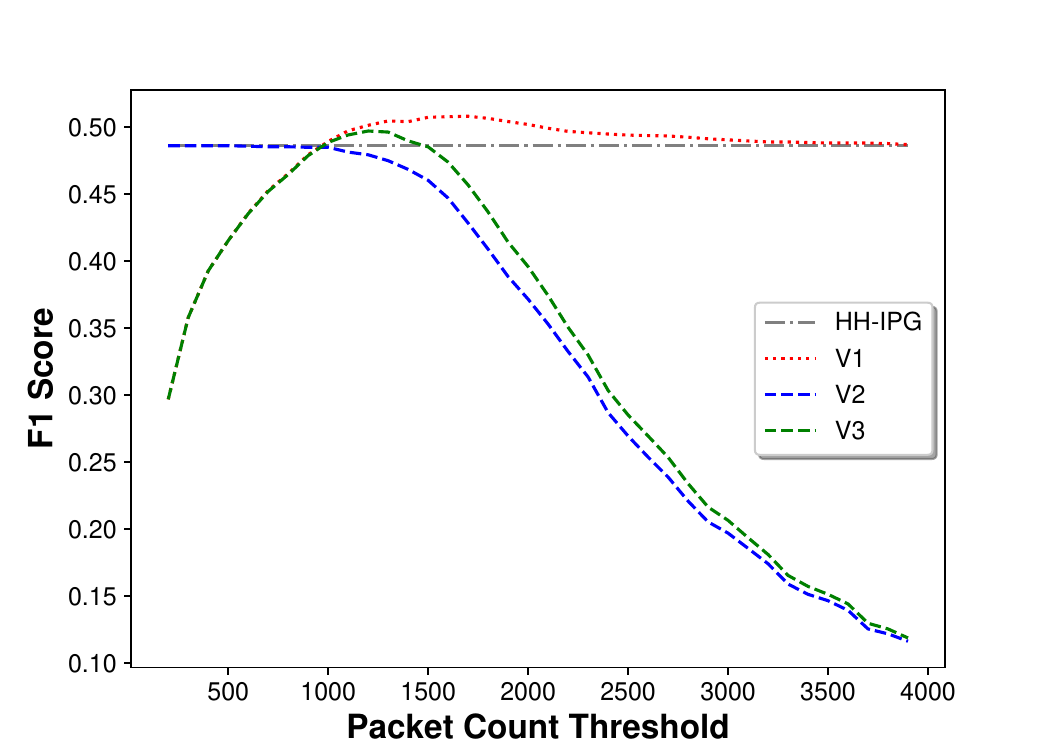}}
\hfill
\subcaptionbox{False Negatives}
{\includegraphics[width=0.23\textwidth]{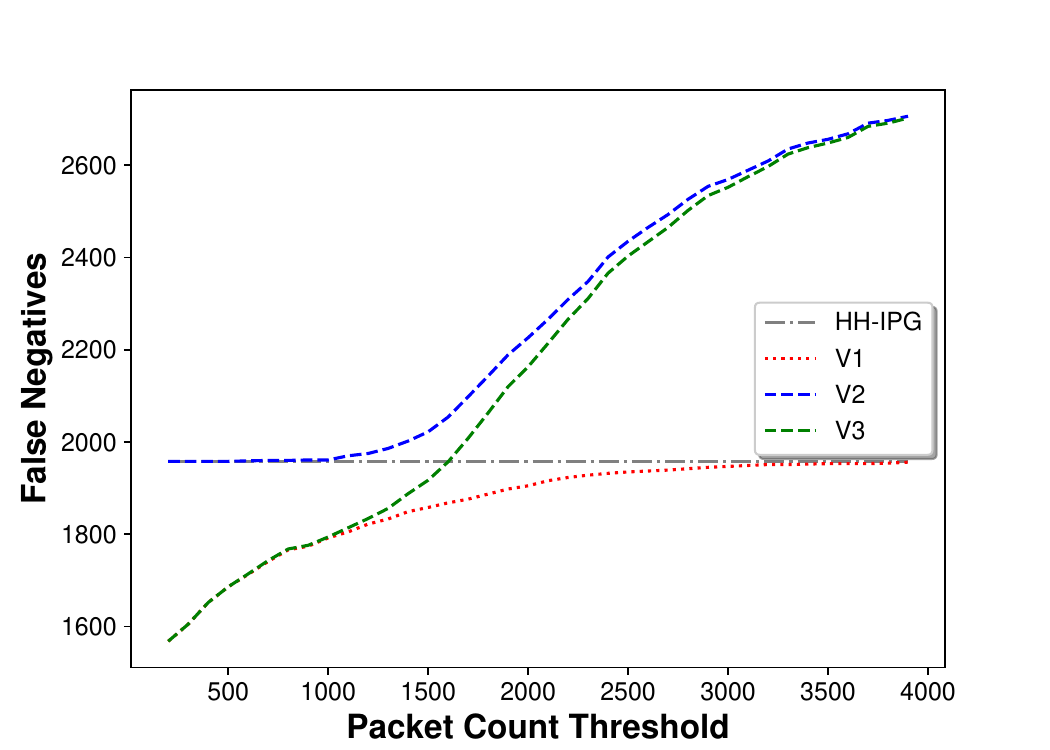}}
\caption{Analysis on CAIDA data trace}  \label{fig:CAIDA2}
\end{figure}

The proposed algorithm performed as expected even on the CAIDA dataset, and few results are provided in this section. The analysis is performed by varying the packet count threshold, for the HH threshold of 1~Mbps and a time window of 5 seconds. The number of heavy-hitters was relatively smaller in the CAIDA dataset compared to MAWI20, and therefore we considered 1~Mbps instead of 10~Mbps as the HH threshold. The result is presented in Fig.~\ref{fig:CAIDA2}. The trace considered consisted of an overall of 2884 heavy hitters, with an average of 216,453 flows per time window. 
We can observe that the resulting pattern is similar to earlier results on the MAWI dataset. The packet count threshold values in range 1k to 1.7k are giving better results. The earlier discussed inferences are hence applicable. The proposed algorithm reduces false negatives by 4.18\% for this trace.

\begin{figure}[hbtp]
\centering
\subcaptionbox{F1 Score} 
{\includegraphics[width=0.23\textwidth]{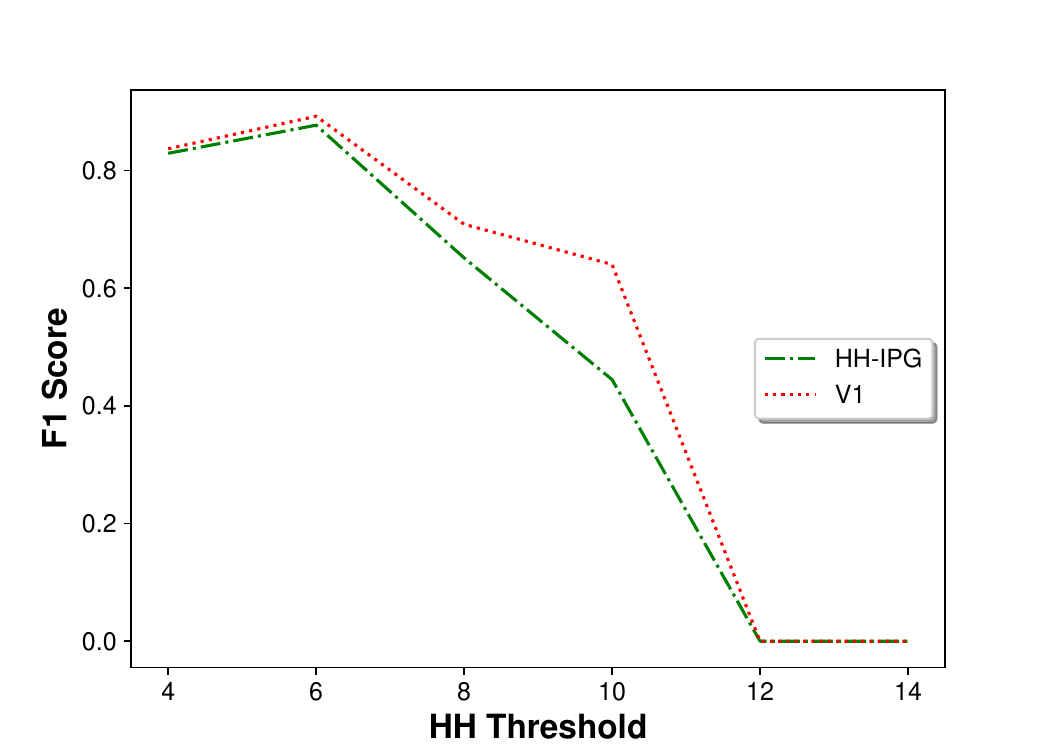}}
\hfill
\subcaptionbox{False Negatives}
{\includegraphics[width=0.23\textwidth]{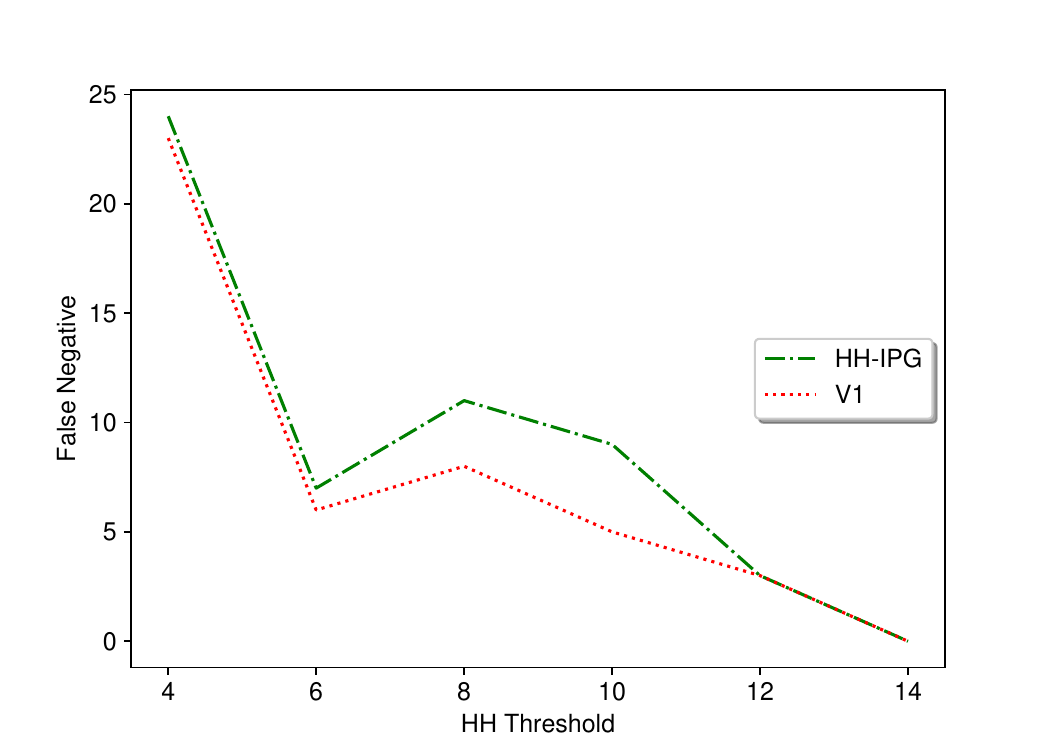}}
\caption{Impact of Heavy Hitter Threshold on CAIDA data trace}  \label{fig:CAIDAhhTh}
\end{figure}

Also, the analysis of the impact of the heavy hitter threshold ($\tau_{th}$) was studied. The result is presented in Fig.~\ref{fig:CAIDAhhTh}, for the time window of 5 seconds. From the results of Fig.~\ref{fig:CAIDA2}, the best packet count threshold of 3100 is chosen for version 1 of algorithm and 1100 for version 2 respectively. As earlier, here also we can observe that the proposed approach results is fewer missed heavy hitters and hence better F1 score. Version 2 and 3 overlaps with the HH-IPG and Version 1 respectively and hence not shown in the results. Therefore, despite choosing packet count threshold by the analysis keeping threshold at a particular point; with the same packet count threshold, the proposed approach yields better results over a range of heavy hitter thresholds. 

\subsection{Using Packet Size} \label{Sec:pSize}

An alternative option for using the packet count (PC) feature is to use the packet byte count or Packet Size (PS) feature. The algorithm using the PS feature is expected to be more accurate than the one with the PC feature for each packet, as the PS value may differ rather than being a constant increment, as in the case of PC. This comes at the cost of an increase in the number of bits required to store the PS value. 

\begin{figure}[hbtp]
\centering
\includegraphics[width=0.45\textwidth]{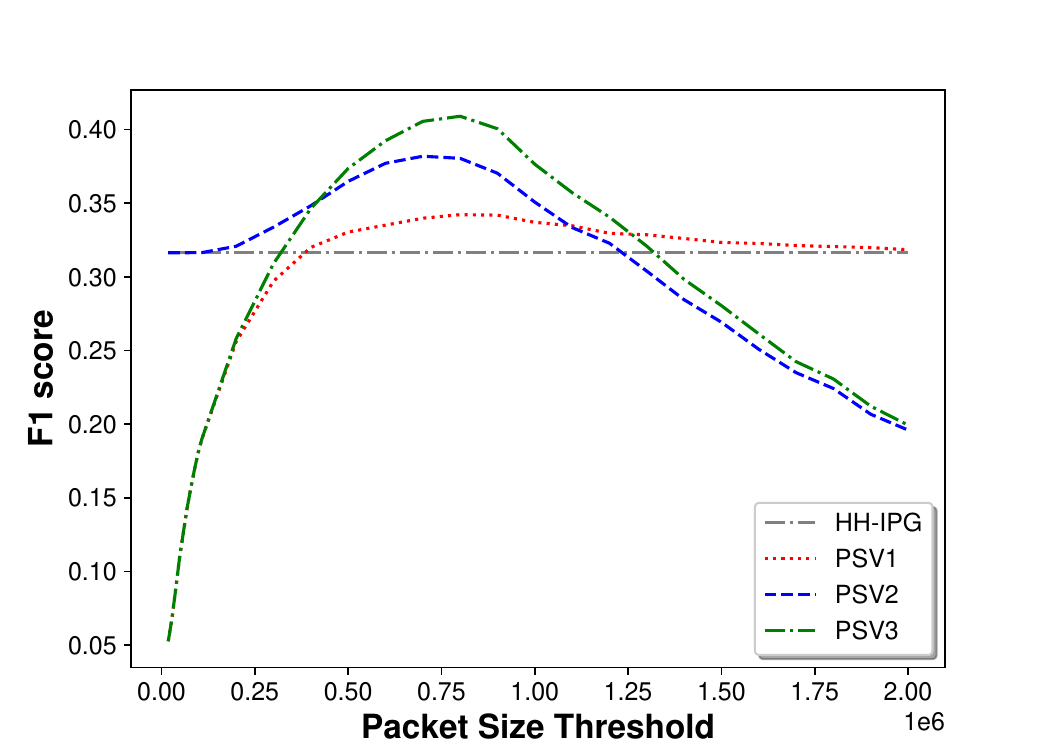}
 \caption{F1 score comparison for packet size feature on CAIDA data trace}
    \label{fig:CAIDA_PS}
\centering
\end{figure}

The analysis is performed by using the PS value in place of the PC, exactly as per the explanations provided for the PC. The result on a subset of CAIDA dataset is provided in Fig.~\ref{fig:CAIDA_PS}. Results shows similar patterns as that of PC, while the improvement of F1 score is better. The usage of PS in place of a PC is more helpful when, in an application network, the packet sizes vary too much. This is because the PS gives better picture of this variance than the PC. 

Quantitatively, for the considered CAIDA trace, there were about 1400 true heavy hitters. The IPG and PC approach improved the F1 score by 2.986\% while the IPG and PS approach further improved it by 5.5521\%. But this comes at the cost of storing values up to 1,000,000 per table entry for packet size, compared with values up to 5000 in the case of packet count. These values are for the thresholds used in algorithm and hence the maximum value that might be stored in that field. 

\section{Conclusions}
This paper presented an enhanced heavy hitter detection algorithm that considered packet count feature in addition to the existing inter-packet gap feature. The objective was to reduce the number of false negatives by considering packet count whenever two flows were mapped to the same table slot in the flow hash table structure. The proposed algorithm was implemented in the P4-enabled Intel Tofino 1 switch and a python based simulator and analyzed using data from public real life datasets. From the performance studies, it was seen that many actual heavy hitter flows have been correctly identified by our modified algorithm, but were missed by the existing IPG-based algorithm. The proposed approach is also shown to be flexible with respect to different parameters.

\subsection*{Funding Declaration}

This work was supported by Ciena Corporation, Ottawa, Canada.

\subsection*{Authors' Contribution Statement}

Adarsha K Shashidhar and Krishna M Sivalingam wrote the main manuscript
text, with design, implementation and performance studies inputs from
Adarsha K Shashidhar, Krishna Sivalingam, Gauravdeep Shami, Marc
Lyonnais and Rodney Wilson.  The software implementation and experiments
were conducted primarily by Adarsha K Shashidhar.

\subsection*{Acknowledgments}

The authors thank Dr. Jim Chen and Fei Yeh of Northwestern University
for providing access to the Tofino switches at the International Center
for Advanced Internet Research (iCAIR), Northwestern University.

\subsection*{Conflict of Interest Statement}

On behalf of all authors, the corresponding author states that there is
no conflict of interest.

\subsection*{Data Availability Statement}

The simulation results that support the findings of this study are
available from the authors but restrictions apply to the public
availability of these data and so are not publicly available. Data are,
however, available from the authors upon reasonable request and with
permission from Ciena Corporation and IIT Madras.

\bibliographystyle{IEEEtran}
\bibliography{reference}


\end{document}